# Thermoelectric properties of iron-based superconductors and parent compounds

*Ilaria Pallecchi, Federico Caglieris, and Marina Putti*
*CNR-SPIN and Università di Genova, Dipartimento di Fisica, Via Dodecaneso 33, 16146 Genova, Italy*

**Abstract**
Herewith, we review the available experimental data of thermoelectric transport properties of iron-based superconductors and parent compounds. We discuss possible physical mechanisms into play in determining the Seebeck effect, from whence one can extract information about Fermi surface reconstruction and Lifshitz transitions, multiband character, coupling of charge carriers with spin excitations and its relevance in the unconventional superconducting pairing mechanism, nematicity, quantum critical fluctuations close to the optimal doping for superconductivity, correlation. Additional information is obtained from the analysis of the Nernst effect, whose enhancement in parent compounds must be related partially to multiband transport and low Fermi level, but mainly to the presence of Dirac cone bands at the Fermi level. In the superconducting compounds, large Nernst effect in the normal state is explained in terms of fluctuating precursors of the spin density wave state, while in the superconducting state it mirrors the usual vortex liquid dissipative regime. A comparison between the phenomenology of thermoelectric behavior of different families of iron-based superconductors and parent compounds allows to evidence the key differences and analogies, thus providing clues on the rich and complex physics of these fascinating unconventional superconductors.

## 1. Introduction

After the great excitement about unconventional superconductivity in cuprates [1], exhibiting unprecedented high values of $T_c$, a new class of unconventional superconductors based on iron was discovered in 2008 [2], exhibiting considerable $T_c$ values up to 58K [3,4]. Given the antagonistic relationship between superconductivity and magnetism, the discovery of these iron-based superconductors has been quite unexpected and has triggered experimental and fundamental studies worldwide. On one hand, their very high upper critical fields, low anisotropy and large $J_c$ values, which are only weakly reduced by magnetic fields at low temperatures, have suggested considerable potential in large scale applications, particularly at high fields. On the other hand, the complete understanding of fundamental mechanisms of unconventional superconductivity is yet to be achieved and extensive investigation of all physical properties should be addressed in this direction. Among key features of superconducting compounds which come into play to determine the mechanisms of superconductivity are the coupling of carriers with boson excitations and the multiband character, relevant to the case of these iron-based materials. These features can be investigated in superconducting compounds as well as in their respective parent compounds through the study of normal state transport properties and in particular of thermoelectric properties. The latter have contributions from diffusion, drag and entropy mechanisms in each band, giving rise to a rich and diverse phenomenology, whose understanding allows to extract precious physical information. Thermoelectric properties, namely Seebeck and Nernst effects, are also ideal tools to probe changes in the Fermi surfaces, changes of scattering mechanisms, presence of different bands with parabolic or linear dispersion crossing the Fermi level. In this work, phenomenological behaviors of thermoelectric properties in iron-based pnictides and chalcogenides are reviewed, presenting both literature and original data. Both common and peculiar aspects of different families are discussed.

## 2. General properties of iron pnictides and chalcogenides

Since the discovery of superconductivity at 26 K in fluorine-doped LaFeAsO [2], many reviews on these compounds have appeared in literature [5,6,7,8,9]. Four main iron-based superconducting families with distinctive crystallographic structures can be identified: the "1111" family with chemical composition $RE$FeAsO ($RE$=rare earth), the "122" family with chemical composition $AFe_2As_2$ or $AFe_2Se_2$ (A=alkaline earth metal), the "111" family represented by LiFeAs (or another alkali metal in the place of Li) and the "11" family with chemical composition FeCh (Ch=chalcogen ion). All the corresponding crystal structures are characterized by square lattices of iron atoms with tetrahedrally coordinated bonds to either phosphorus, arsenic, selenium or tellurium anions that are staggered above and below the iron lattice. These slabs are either simply stacked together, as in FeSe, or separated by spacer layers using alkali (for example, Li), alkaline-earth (for example, Ba), rare-earth oxide/fluoride (for example, LaO or SrF). The geometry of the $FeAs_4$ tetrahedra, and specifically the As-Fe-As bond angle, play a crucial role in determining the electronic, magnetic and superconducting properties of these systems.

Long-range magnetic order also shares a similar pattern in all of the iron-based parent compound systems, with an arrangement consisting of spins ferromagnetically arranged along one chain of nearest neighbours within the iron lattice plane, and antiferromagnetically arranged along the other direction.

Common features are easily identified also in the electronic band structures of iron-based superconductors. The dominant contribution to the electronic density of states at the Fermi level derives from metallic bonding of the iron $d$-electron orbitals, which form a Fermi surface of at least four quasi-2D electron and hole cylinders. These consist of two hole pockets at the Brillouin zone centre and two electron pockets at $(0,\pm\pi)$ and $(\pm\pi,0)$ in the tetragonal unit cell. A fifth hole band may be also present at $(0,\pm\pi)$, depending on structural and compositional details.

A spin density wave (SDW) instability arises from the nesting of two Fermi surface pockets by a large $Q=(\pi,\pi)$ vector that is commensurate with the structure. This vector corresponds to the magnetic ordering vector measured throughout the FeAs-based parent compounds as well as that for magnetic fluctuations in the related superconducting compounds. Experimental evidence for $(\pi,\pi)$ Fermi surface nesting across most of iron-based compounds has been found. In facts, a notable exception is $A_xFe_{2-y}Se_2$, where A is an alkali element [10].

Since the earliest stages of research on these materials, the unconventional nature of the pairing mechanism has been pointed out and although the mediator of pairing is yet unidentified, it is widely believed that is should be attributed to magnetic spin fluctuations. In particular, the spin-mediated mechanism assumes an exchange of antiferromagnetic fluctuations between the hole and electron pockets connected by the antiferromagnetic wave-vector $Q=(\pi,\pi)$. In this picture, magnetism must be suppressed, either by pressure or doping, before optimal bulk-phase superconductivity appears. The resulting multiband pairing gap symmetry is indicated to be s-wave, with a sign change of the order parameter in different sheets of the Fermi surface ($s\pm$ symmetry) by most of the experimental evidences and theoretical predictions.

In the 11 family, magnetism appears to be peculiar. Indeed, the nesting wave-vector for the electron and hole Fermi surface pockets is also $(\pi,\pi)$, but the experimental in-plane magnetic propagation vector is $(\pi,0)$ in $Fe_{1+y}Te$, suggesting that in this compound antiferromagnetism does not originate from the Fermi surface nesting of itinerant charges, but rather from local magnetic moments. However, with increasing x in the $Fe_{1+y}(Te_{1-x}Se_x)$ system, $(\pi,\pi)$ spin fluctuations increasingly replace $(\pi,0)$ spin fluctuations and correspondingly superconductivity appears, reconciling the behavior of the 11 family with that of other iron-based families [11,12].

In the phase diagrams of the main iron-based families, the undoped parent compounds undergo a tetragonal-orthorhombic structural transformation upon cooling at the temperature $T_s$, closely followed by a magnetic transition at $T_N$ (also indicated as $T_{SDW}$ throughout this paper, in order to comply with the notation of the cited publications in each case, when antiferromagnetism is associated to the SDW order). In most cases $T_N$ and $T_s$ almost coincide within few K. With increasing doping, the phase diagrams are characterized by the competition between magnetic and superconducting orders. The magnetic order observed in the parent compounds below $T_N$ disappears at doping levels of one to few percent. A superconducting ground state appears with increasing doping, below a dome shaped region of the phase diagram, with maximum superconducting $T_c$ at optimal doping. In some cases, the magnetic and superconducting phases coexist in some portions of the phase diagram, with phase separation at nanoscale level.

The signatures of structural/magnetic transition are seen in most normal state properties, for example resistivity, which in general is weakly temperature dependent at high temperature and undergoes abrupt drop and change of curvature in correspondence of the transition, followed by metallic behavior at lower temperatures.

## 3. Physical mechanisms of thermoelectric transport
### 3.1 Phenomenology of thermoelectric effects

Seebeck and Nernst effects are the thermoelectric properties originated in open-circuit conditions from thermal diffusion of charge carriers upon application of a temperature gradient. If an external field is applied perpendicular to the direction of charge carrier diffusion, the deflecting effect of the Lorentz force must be also taken into account. Possible situations are schematically sketched in Figure 1. The average particle velocity is larger at the hot side as compared to the cold side, resulting in a net charge diffusion along the direction of the thermal gradient (panel a)). In open-circuit stationary conditions, a longitudinal voltage drop (Seebeck voltage) builds up and drives a drift charge current that balances the diffusive charge current (panel b)). As a result, the net charge flow is null. In this naïve picture, in presence of a perpendicular magnetic field the net effect of the Lorentz force is null and the transverse voltage drop which is proportional to the Nernst voltage would be zero. In the case that charge carriers of different signs, holes and electrons, contribute to transport (panel c)), thermal diffusion of carriers occurs even with zero net charge flow, and for nearly equal electron and hole densities, negligible longitudinal voltage appears. On the other hand, holes and electrons moving in the same direction are deviated in opposite directions by a perpendicular magnetic field, giving rise to a non-vanishing Nernst voltage.

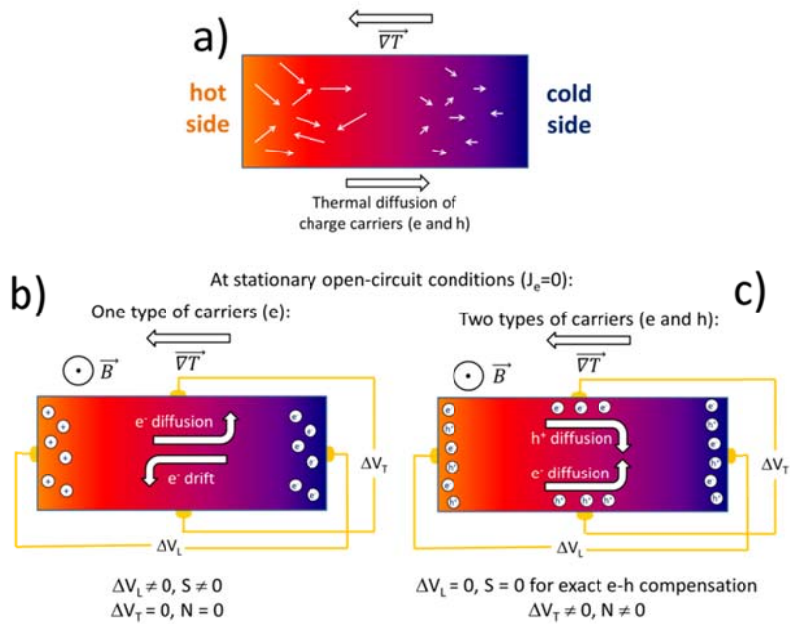

**Figure 1:** Sketch of the experimental configuration for thermoelectric measurements, with applied thermal gradient ($\overline{\nabla} T$) and magnetic field ($\overline{B}$) and measured longitudinal ($\Delta V_L$) and transverse ($\Delta V_T$) voltage drops. The effect of diffusion and drifts currents of charge carriers of different signs as well as of the Lorentz force is shown and described in the text.

Such minimal description can be formalized in a semiclassical model, where the transport equations are written in terms of electric, thermal and Peltier conductivity tensors, σ, κ and α, which relate charge current $J_e$ and heat current $J_h$ to electric field $\overline{E}$ and thermal gradient $\overline{\nabla T}$ vectors. In absence of net charge current ($J_e=0$), the following equation holds:

$$\overline{E} = \sigma^{-1} \cdot \alpha \cdot \overline{\nabla T} \tag{1}$$

Within this tensorial relationship, the Seebeck and Nernst effects describe the longitudinal and transverse electric fields generated by a longitudinal thermal gradient, respectively. If we assume an isotropic medium and indicate by x and y the longitudinal and transverse directions, the S and N signals are defined as:

$$S = \frac{E_x}{\nabla_x T} = \frac{\sigma_{xx}\alpha_{xx} - \sigma_{xy}\alpha_{xy}}{\sigma_{xx}^2 + \sigma_{xy}^2} \approx \frac{\alpha_{xx}}{\sigma_{xx}} \tag{2}$$

$$N = -\frac{E_y}{\nabla_x T} = \frac{\sigma_{xx}\alpha_{xy} - \sigma_{xy}\alpha_{xx}}{\sigma_{xx}^2 + \sigma_{xy}^2} \tag{3}$$

where in eq. (2) it is taken into account that $\sigma_{xy} \ll \sigma_{xx}$ and $\alpha_{xy} \ll \alpha_{xx}$. Note that eq. (3) is also often expressed without the minus sign. However, throughout this paper, we choose to comply with the recent sign convention that associates a positive Nernst coefficient to the motion of superconducting vortices, thus requiring the minus sign.

**3.2 Mechanisms contributing to Seebeck effect**
The elements of the conductivity tensors can be explicitly written using Boltzmann transport equations. Cutler and Mott [13] derived the diffusive (i.e. originating from thermal diffusion of charge carriers) Seebeck coefficient in this formalism as:

$$S = -\frac{k_B}{q}\frac{1}{\sigma}\int_0^\infty \sigma(E)\left(\frac{E-E_F}{k_B T}\right)\left(\frac{\partial f_0(E)}{\partial E}\right)dE \tag{4}$$

where q is the magnitude of the electron charge, $E_F$ is the Fermi energy, $\sigma$ the electric conductivity, $k_B$ the Boltzmann constant and $f_0$ the equilibrium Fermi distribution. In systems with degenerate Fermi statistics eq. (4) can be simplified to the well-known Mott expression:

$$S = \pm\frac{\pi^2}{3}\frac{k_B^2 T}{q}\left[\frac{\partial \ln\sigma(E)}{\partial E}\right]_{E=E_F} \tag{5}$$

which is generally valid regardless the conduction mechanism, either through band or localized states. In Eq. (5) S has the sign of the charge carriers. Eq. (5) can be used either in a single band picture or for each band of a multiband picture. Furthermore, if the conductivity is expressed as $\sigma = qn\mu = q^2 n\tau/m_{eff}$, where n is the carrier density, $\mu$ the mobility, $\tau$ the scattering time and $m_{eff}$ the effective mass, the Mott relationship eq. (5) can be written as a sum of several contributions, related to the energy dependence of n, $\tau$ and $m_{eff}$. As a consequence, the diffusive Seebeck coefficient depends weakly on disorder, which is described by a logarithmic additive term proportional to $\left[\frac{\partial \ln(\tau(E))}{\partial E}\right]_{E=E_F}$. The details of the band structures are approximately described by the additive term proportional to $\left[\frac{\partial \ln(m_{eff}(E))}{\partial E}\right]_{E=E_F}$. However, in the most simplified situation only the energy dependence of n gives non negligible contribution, so that eq. (5) can be written as:

$$S = \pm\frac{\pi^2}{3}\frac{1}{q}\frac{k_B^2 T}{E_F} \tag{6}$$

Hence, to the leading order, the diffusive Seebeck coefficient for a single band depends linearly on the temperature. An evident dependence on the magnetic field may occur either due to the dependence of the density of states on H in the quantum regime or due to magnetic field related scattering mechanisms.

If multiple bands contribute to transport, eq. (5) applies to each band and the total S result from the parallel contribution of all the bands, as the sum of the Seebeck coefficients of each band $S_j$ weighed by the respective electrical conductivities $\sigma_j$, as:

$$S = \frac{\sum_j \sigma_j S_j}{\sum_j \sigma_j} \qquad (7)$$

In a multiband picture the overall temperature dependence of the diffusive S may exhibit very different behaviors, determined by temperature dependent charge densities and mobilities of each band.

It is worth mentioning a limiting case of diffusive Seebeck, occurring in narrow-banded materials with strong on-site Coulomb repulsion $U \gg k_B T$. In such materials at sufficiently high temperature, the kinetic terms contributing to S can be neglected, so that the value of S, representing the entropy carried by each charge carrier, can be calculated by simple combinatorial arguments. Indicating by $n_{site}$ the carrier density per atomic site, the single band S is described by the Heikes law [14,15,16] predicting constant temperature dependence:

$$S(T \to \infty) = \pm \frac{k_B}{q} \ln\left(2 \frac{1 - n_{site}}{n_{site}}\right) \qquad (8)$$

Besides the diffusive mechanisms, Seebeck voltage can be also generated by the so called drag mechanisms. The phonon drag contribution ($S_{ph}$) is created because phonons diffuse along the direction of the thermal gradient just like charge carriers. It is due to the momentum transfer between the system of phonons and the system of charge carriers and it is observed in the temperature regime where phonons thermalize by scattering preferentially with charge carriers. A phenomenological expression of the phonon drag contribution is given by [17]:

$$S_{ph} = \pm 3 k_B \left(\frac{\alpha_{ph}(T)}{q}\right)\left(\frac{T}{\theta}\right)^3 \int_0^{\theta/T} \frac{x^4 e^{-x}}{(1 - e^{-x})^2} dx \qquad (9)$$

where $\theta$ is the Debye temperature, the integral represents the phonon specific heat and $\alpha_{ph}$ is the effective drag parameter, averaged over the phonon spectrum (where the average over the phonon spectrum is carried out in such a way that $\alpha_{ph}(T)$ can be taken out of the integral). This parameter, whose value is in the range $0 < \alpha_{ph} < 1$, takes into account the phonon-electron interaction effectiveness and can be expressed as:

$$\alpha_{ph} \approx \frac{\tau_{phe}^{-1}}{\tau_{phx}^{-1} + \tau_{phe}^{-1}} \qquad (10)$$

where $\tau_{phe}^{-1}$ is the phonon scattering rate by electrons and $\tau_{phx}^{-1}$ is the phonon scattering rate by any mechanism other than by electrons (phonon-grain boundary, phonon-defect, phonon-phonon). The $S_{ph}$ contribution to S is easily identified as it is peaked at a typical temperature $\theta/5 - \theta/4$. Indeed, at low temperatures it grows according to the temperature excitation of phonon modes, namely $\sim T^3$ if the temperature dependence of the $\alpha_{ph}$ parameter can be neglected in this low temperature regime, while at larger temperatures approaching $\theta$, the density of excited phonons increases and the phonons are mainly thermalized by scattering preferentially with other phonons rather than with electrons, making $S_{ph}$ vanish ($\tau_{phx}^{-1} \gg \tau_{phe}^{-1}$, where $\tau_{phx}$ describes the phonon-phonon scattering, so that $\alpha_{ph} \ll 1$ in eq(10)). Form eq. (9) and (10) it can be seen that $S_{ph}$ is easily suppressed by disorder

($\tau_{phx}^{-1} >> \tau_{phe}^{-1}$, where $\tau_{phx}$ describes the phonon-defect scattering, so that $\alpha_{ph} << 1$ in eq(10)). On the contrary, the magnetic field dependence of $S_{ph}$ is expected to be weak.

Any system of bosons that exchanges momentum with the system of charge carriers introduces in principle a drag contribution to the Seebeck effect in a characteristic temperature range. Relevant to the case of iron pnictides and chalcogenides is the drag contribution of the antiferromagnetic (AFM) spin fluctuations or magnons. A detailed modelling of AFM magnon drag in these systems is presented in ref. [18]. The magnon drag contribution to the Seebeck effect and its temperature dependence are pretty analogous to the phonon drag contribution, in that it is proportional to the magnon specific heat $C_m$ and to the scattering parameter $\alpha_m \approx \frac{\tau_{me}^{-1}}{\tau_{mx}^{-1} + \tau_{me}^{-1}}$ where $\tau_{me}^{-1}$ is the magnon-electron scattering rate and $\tau_{mx}^{-1}$ the magnon scattering rate with any other relaxing mechanism (magnon-grain boundary, magnon-defect, magnon-phonon, magnon-magnon). This suggests that the magnon drag Seebeck is easily suppressed by the disorder that is effective in obstructing the propagation of spin waves, similarly to the sensitivity to disorder of the phonon drag Seebeck. Moreover, like the phonon drag Seebeck that mirrors the temperature dependence of the phonon specific heat at low temperature, also the magnon drag Seebeck mirrors the temperature dependence of the magnon specific heat, at least in the regimes where the temperature dependence of the drag parameter $\alpha_m$ can be neglected. Indeed, in the limit $\Delta << k_B T << T_N$ ($\Delta$ is the gap of the magnum spectrum) the AFM magnon drag Seebeck is expected to be proportional to $T^3$, at odds with the ferromagnetic (FM) magnon drag Seebeck that obeys $T^{3/2}$ dependence [19], while in the opposite limit $k_B T << \Delta$ the AFM magnon drag Seebeck should exhibit an activated temperature behavior $\sim T^{1/2} \exp(-\Delta/k_B T)$, describing excitation of magnons in the gapped spectrum. In ref. [18] it is shown that the magnetic field dependence of the magnon drag Seebeck is expected to be significantly stronger than that of the phonon drag Seebeck. Namely, the magnon drag Seebeck is increased in magnitude as a function of the longitudinal field $B_\parallel$ (i.e. oriented along the spin direction), especially at low temperatures. Remarkably, the AFM magnon drag is a growing function of the magnetic field, oppositely to the FM magnon drag that is suppressed by field [19]. Physically, a magnetic field helps (contrasts) the creation of magnons in the spin sublattice oriented antiparallel (parallel) to the field. In the limit $k_B T << \Delta$, the AFM magnon drag obeys an approximate universal scaling behavior as a function of temperature T and longitudinal applied field $B_\parallel$, namely it depends only on the ratio $B_\parallel/T$, with a functional form that depends on the details of the magnon spectrum.

**3.3 Mechanisms contributing to Nernst effect**

The Nernst effect is the magneto-thermoelectric effect, defined as the appearance of a transverse electric field $E_y$ in response to a temperature gradient $\overline{\nabla} T \parallel x$, in the presence of a perpendicular magnetic field $\overline{B} \parallel z$ ($\overline{E} \perp \overline{B} \perp \overline{\nabla} T$) and under open circuit conditions. While the Nernst signal is given by eq. (3) $N = -E_y / \nabla_x T$, the Nernst coefficient $\nu$ is defined as $\nu = N/B$ in the range where N is linear in B. Within the Boltzmann theory, in a single band picture, the Nernst coefficient can be expressed as:

$$\nu = \frac{N}{B} = -\frac{\pi^2}{3} \frac{k_B^2 T}{qB} \frac{\partial \tan\theta_H}{\partial E}\bigg|_{E=E_F} = -\frac{\pi^2}{3} \frac{k_B^2 T}{m_{eff}} \frac{\partial \tau}{\partial E}\bigg|_{E=E_F} \qquad (11)$$

where $\tan(\theta_H) = qB\tau/m_{eff}$ is the Hall angle. Eq. (11) evidences that if the Hall angle does not depend on the energy, $\nu$ vanishes. This situation can be equivalently formulated in terms of the so-called Sondheimer cancellation [20]:

$$\nu = \left(\frac{\alpha_{xy}}{\sigma} - S \tan\theta_H\right)\frac{1}{B} \qquad (12)$$

Eqs. (11) and (12) explain why the Nernst effect is usually small in conventional metals. Either multiband effects or energy dependent Hall angle (equivalently energy dependent mobility μ) make the Sondheimer cancellation no longer valid. Eq. (11) can be further developed if the energy dependence of the Hall angle is explicitly expressed. In ref. [21] a linearized energy dependence of the Hall angle in the vicinity of $E_F$ is assumed, yielding $\nu = -\frac{\pi^2}{3}\frac{k_B T}{q}\frac{\mu}{T_F}$. On the other hand, in the case of elastic impurity scattering, $\tau \sim E^{-1/2}$ is customarily assumed, corresponding to an energy independent mean free path. In this case eq. (11) becomes:

$$\nu = \frac{\pi^2}{6}\frac{k_B T}{q}\frac{\mu}{T_F} \tag{13}$$

Eq. (13) shows that a large electron mobility μ and a small Fermi energy yield large Nernst. For example, in presence of Fermi surface reconstruction arising in a spin density wave ordered state, Sondheimer cancellation may be severely violated and enhancement of the Nernst effect associated to Fermi surface pockets is possible.

As for multiband effects, the Nernst effect can be modestly enhanced by the existence of two types of carriers in the system, in particular the ambipolar Nernst signal is maximal when the bands are exactly compensated. This can be understood by thinking that charge carriers of opposite sign are driven by the thermal gradient along the same direction and thus are deflected along opposite directions by the magnetic field (see Figure 1c)), at odds with the electric transport counterpart of the Nernst effect, i.e. the Hall effect, where charge carriers of opposite sign are driven by the electric field along opposite directions and thus are deflected along the same direction by the magnetic field. Hence, in compensated compounds the Hall resistance vanishes, whereas the Nernst effect is magnified. The expression of N in a two band formulation is [21]:

$$N = \frac{(\alpha_{xy}^+ + \alpha_{xy}^-)(\sigma_{xx}^+ + \sigma_{xx}^-) - (\alpha_{xx}^+ + \alpha_{xx}^-)(\sigma_{xy}^+ + \sigma_{xy}^-)}{(\sigma_{xx}^+ + \sigma_{xx}^-)^2 + (\sigma_{xy}^+ + \sigma_{xy}^-)^2} \tag{14}$$

where the superscripts + and – refer to hole like and electron line carriers.

As for the magnetic field dependence of the Nernst effect, Eq. (11) can be rearranged [22] to evidence that the magnitude of the Nernst effect is controlled by the product $k_B T\tau/\hbar$. This can be small, e.g., when the scattering is strong or at low temperatures in the impurity dominated regime, or large, e.g., in clean systems with only moderate inelastic scattering. In this latter regime, the Boltzmann theory predicts that the range of magnetic fields over which N is linear with B diminishes, with the crossover to N~1/B taking place at $\omega_c\tau \approx 1$ ($\omega_c = qB/m_{eff}$ cyclotron frequency).

Much larger enhancement of N than that related violation of the Sondheimer cancellation due to multiband effects or energy dependent Hall angle may occur in case of linear dispersion of one band [22]. This situation is indeed relevant to iron-pnictides, as discussed theoretically [23] and demonstrated experimentally [24]. The different energy dispersion relationship and scattering rates of Dirac fermions and conventional electrons may lead to anomalous temperature dependence of the transport coefficients, in particular of Nernst coefficient. In ref. [22], the expression for the Nernst coefficient in a two-dimensional system with a band having Dirac dispersion is extracted in the limit $\tilde{\mu}\tau > \hbar$, where $\tilde{\mu}$ is the chemical potential with respect to the Dirac cone vertex:

$$\nu = -\frac{\pi^2}{3}\frac{k_B^2 T v_{dc}^2 \tau}{\tilde{\mu}^2} \tag{15}$$

Here $v_{dc}$ is the velocity $(1/\hbar)dE/dk$ of Dirac cones. From eq. (15) it comes out that the Nernst effect is proportional to the inverse squared chemical potential and thus it is highly enhanced if the Fermi level lies close to the Dirac cone vertex, forming tiny Fermi surface pockets, as may occur in pnictide parent compounds [24]. When the condition $\tilde{\mu}\tau > \hbar$ does not hold because the Fermi level lies very close to the Dirac cone vertex, the Nernst effect is expected to be positive and proportional to the third power of the scattering time [22], which can be large in clean systems:

$$\nu = C \frac{k_B^2 T v_{dc}^2 \tau^3}{\hbar^2} \qquad (16)$$

where C is a numerical constant. However eq. (16) may define an initial large linear slope of N as a function of the magnetic field, rapidly followed at $\omega_c \cdot \tau \approx 1$ by a saturation as a function of magnetic field to the Boltzmann value $N \sim k_B^2 T \tau / q\hbar$.

Up to now transport in the normal state has been considered, however superconducting state introduces additional mechanisms that contribute to the Nernst effect [25]. In the presence of a magnetic field $\overline{B} \| z$, the temperature gradient $\overline{\nabla} T \| x$ drives flux vortices along the x-axis direction with velocity $\overline{v}$. Moving vortices produce a transverse electric field $E_y = \overline{B} \times \overline{v}$, which is called vortex Nernst effect. In fact, thermoelectric effects related to vortices are associated solely to the Nernst effect and not to the Seebeck effect. The latter is associated with entropy and heat transport parallel to the induced electric field. This longitudinal entropy flow is not carried by the normal excitations in the vortex cores since these excitations move with the vortices perpendicular to the induced electric field. Instead, the longitudinal entropy transport has to be attributed to other excitations, for example to quasi-particles excited over the energy gap as in high-$T_c$ cuprates [26]. Indeed, the Seebeck effect in the mixed state as a function of temperature at different fields exhibits a different behavior as compared to the behavior of the mixed state longitudinal resistivity dominated by thermally activated vortex motion [27].

Finally, as widely observed in the case of high-$T_c$ cuprates [28], superconducting fluctuations above $T_c$ may contribute to a sizeable Nernst effect [29]. As derived in ref. [30], thermoelectric coefficients can be expressed in terms of the temperature derivative of the chemical potential $\partial\tilde{\mu}/\partial T$, via the respective relationships with the conductivity tensor, as an alternative to the classical kinetic approach, where transport equations are formulated and solved. It turns out that the Nernst effect is composed of two terms, the first one is governed by the temperature dependence of the chemical potential, while the second is related to magnetization currents and is relevant in case of superconducting fluctuations above $T_c$.

**4. Seebeck effect of parent compounds of iron pnictides and chalcogenides**
**4.1 Seebeck effect of 1111 parent compounds**
The 1111 parent compounds undergo crystallographic transitions from the tetragonal high-temperature structure to the orthorhombic low-temperature structure around 130-150 K. In the low-temperature phase, stripe antiferromagnetic ordering (with spins ordering ferromagnetically in one direction and antiferromagnetically in the other direction in real space) of small moments (< 1$\mu_B$, $\mu_B$ Bohr magneton) on the iron sites occurs, forming a commensurate SDW. A further magnetic transition may occur if the rare-earth (*RE* in the chemical formula) moments order antiferromagnetically and this occurs at different temperatures ranging from 14K to 2K, depending on the rare-earth itself. When these materials are doped, the structural and magnetic transitions at ~150 K are suppressed and superconductivity emerges. Transport and magnetic properties are affected strongly by the structural and magnetic transitions, suggesting that they are associated to significant changes in the band structure and/or carrier mobilities.

In Figure 2, the temperature dependent Seebeck coefficient curves measured in a series of pnictide parent 1111 compounds, having different chemical composition *RE*FeAsO (*RE*=La, Ce, Pr, Nd, Sm) are shown. At high temperature all the curves are negative and increase in absolute value with decreasing temperature. Around the structural/magnetic transition, which varies between 130K and 145K among these compounds, all the curves undergo an abrupt change. Below the transition, the curves follow different behaviors, before eventually vanishing in the limit of zero temperature. In some cases, the Seebeck curves change in sign, becoming positive at low temperature, while other curves are negative. However, even in this low temperature regime, a general feature is observed, namely the presence of a broad bump, responsible for a minimum of S around 50K. The most remarkable feature, noted in ref. [18], is that the low temperature behavior (T<$T_{SDW}$) seems to change from sample to sample, even with the same *RE,* whereas the dependence on the rare earth *RE* appears to be weaker. By converse, the high temperature (T>$T_{SDW}$) behavior is largely consistent between different *RE*FeAsO and samples, assuming values between -6 and -19 µV/K at 300K.

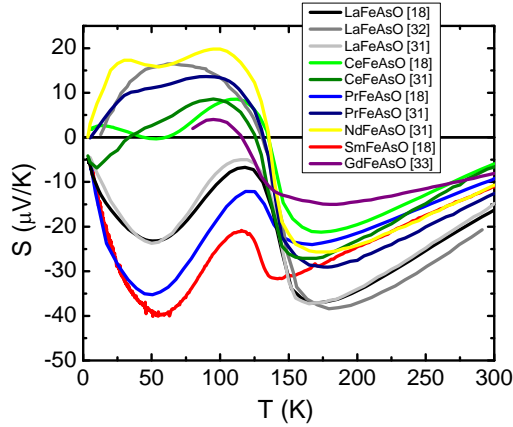

**Figure 2:** Seebeck coefficient curves of *RE*FeAsO (*RE*=La, Ce, Pr, Nd, Sm, Gd) polycrystals taken from ref. [18,31,32,33].

We first focus on the sharp change of behavior at $T_{SDW}$. The S curves of these parent compounds around the transition are pretty insensitive to applied magnetic field [34]. In ref. [34], the thermopower is expressed as the sum of two contributions, namely the usual diffusive one plus one arising from temperature dependence of the chemical potential $S_{\tilde{\mu}}$. The latter is a direct measure of the electron entropy and may become non negligible near transitions. As the electronic specific heat $C_e$ is also a measure of the electron entropy, $C_e$ should be linearly related to the temperature derivative of $S_{\tilde{\mu}}$ multiplied by the temperature, $C_e \propto T \frac{dS_{\tilde{\mu}}}{dT}$. The dominance of the $S_{\tilde{\mu}}$ term in the temperature dependence of the measured S close to the transition is indeed verified in NdFeAsO and SmFeAsO [34] by the linear proportionality $C_e \propto T \cdot dS/dT$. The sharp change of behavior at $T_{SDW}$ could be alternatively attributed to a dramatic change in the charge carrier scattering mechanism or scattering time at the transition [35,36], according to eq. (5), or else to a multiband mechanism. For example, the change of S behavior while lowering the temperature across the transition could be reproduced assuming an abrupt reduction of the contribution of the electron bands to the total S, or similarly an abrupt enhancement of the contribution of the hole bands. It is likely that more than one single mechanism is active across the transition.

We now focus on the low temperature regime. We consider the LaFeAsO curves in Figure 2, taken from Caglieris et al. [18], McGuire et al. [31] and Kondrat et al. [32], respectively. Above $T_{SDW}$ the three curves nearly overlap, while for T<$T_{SDW}$ the curves exhibit sharply different behaviors, being McGuire's and Caglieris' data characterized by a negative bump, and Kondrat's data by a rounded positive maximum. In ref. [18], the resistivity curves of the same samples are compared and it is found that the sample of ref. [32] shows a more significant resistivity upturn, suggesting that carrier localization by disorder may be simultaneously responsible for both the resistivity upturn and for the absence of the low temperature negative bump in the Seebeck effect. This low temperature behavior of thermopower can be further investigated by exploring the effect of an applied magnetic field, as shown in Figure 3 for the sample of Caglieris et al. [18]. Above $T_{SDW}$ the dependence of S on magnetic field is negligible, while it is maximum, larger than 20%, in correspondence of the low temperature bump. Similar field enhancement of the bump feature is observed also in SmFeAsO [34], NdFeAsO [34] and LaFeAsO [37], whereas the Seebeck effect of the sample from ref. [32], which shows no bump, does not depend on the field [38]. Thus it can be concluded, that the Seebeck bump is magnetic field dependent, more specifically enhanced by an applied magnetic field and easily suppressed by disorder. From these clues, the negative bump is identified as magnon drag Seebeck contribution, superimposed to the ever present diffusive Seebeck term. As a consequence of the multiband character of the compound, the diffusive Seebeck contribution shows a non trivial temperature dependence, which turns out to be pretty similar to the Seebeck curve of the Kondrat's sample [32]. This allows to obtain the net magnon drag Seebeck contribution by subtraction and explore its temperature and field dependences. It turns out that it scales as a function of the ratio B/T, which is a consequence of thermal activation of magnons, as predicted by the model developed in ref. [18]. It can be noted that the magnon drag scenario described so far is not in contrast with the lack of $T^3$ dependence of S, typical of drag contributions at low temperature, as pointed out in ref. [27,34]. Indeed, as discussed in section 3.2, if the magnon gap is not negligible as compared to the temperature, which is indeed the present case, the temperature dependence of S is not expected to replicate the $T^3$ dependence of the magnon specific heat. In ref. [18], it is further shown that the specific heat data of LaFeAsO exhibit negligible field dependence, indicating that magnon density is much smaller than phonon density. Even this result does not contradict the magnon drag picture, as the field independence of the specific heat combined with the strong field dependence of the Seebeck effect are reconciled assuming that the drag parameter for the magnons $\alpha_m$ is very large as compared to its phonon counterpart $\alpha_{ph}$, namely $\alpha_{ph} \ll \alpha_m$. This is an evidence that charge carriers are much more coupled with magnons than with phonons. Such coupling of electrons with AFM spin fluctuations in these parent compounds has crucial implications on the pairing mechanism of the related superconducting compounds.

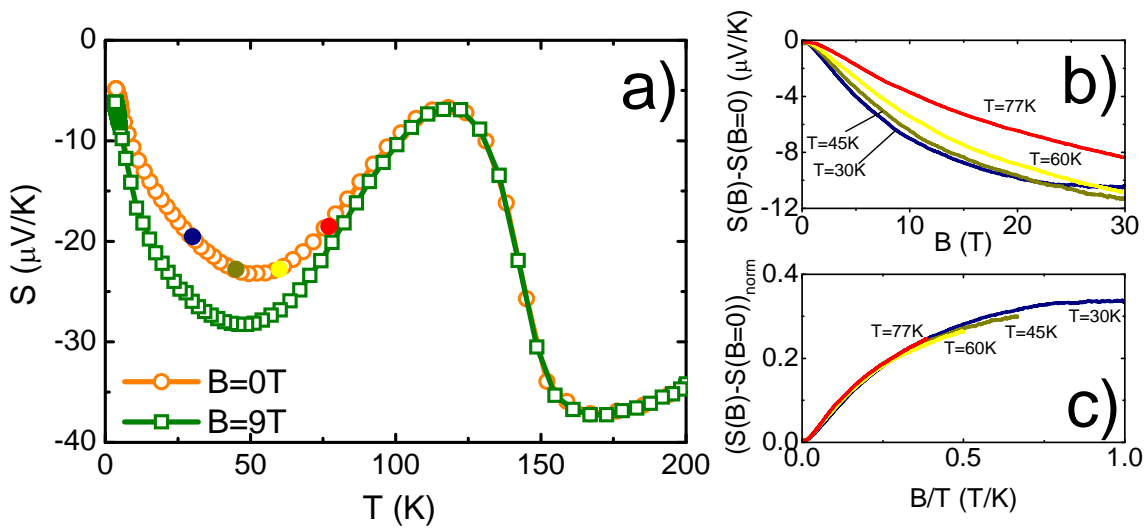

**Figure 3:** a) Seebeck coefficient curves of LaFeAsO from ref. [18] measured at zero and 9T. b) S curves versus the magnetic field up to 30T measured at fixed temperatures T= 30, 45, 60, 77 K. c) Normalized magnon drag contribution to the Seebeck effect extracted from the experimental S curves of panel b) and plotted as a function of B/T.

The investigation of chemical substitution on the parent compounds, even when not yielding superconductivity, gives information on the effects of band filling, band structure changes and disorder on the ground state and on transport properties of the compounds. Since the Seebeck effect is a sensitive probe of band parameters and scattering mechanisms, it is interesting to consider the behavior of Seebeck curves in some non-superconducting doped 1111 compounds. In PrFe$_{1-x}$Ru$_x$AsO, studied in ref. [39], the SDW is completely suppressed upon Ru substitution and the system is made increasingly metallic, as a consequence of the increased bandwidth and hybridization with $p$ orbitals with Ru doping. A similar trend is also found in LaFe$_{1-x}$Ru$_x$AsO. Accordingly, in Seebeck curves shown in the left panel of Figure 4, the feature associated to the SDW transition is broadened and shifted to lower temperatures with increasing doping. At odds with Co and Ni substitution on the Fe site (see section 5.1), no evidence of superconductivity above 2 K is found, which is possibly related to the distortion of the tetrahedral coordination environment of the transition metal site that increases as the Ru concentration increases. As also seen in the left panel of Figure 4, a strong positive enhancement of the low temperature Seebeck coefficient is observed at low doping $x = 0.10$ for both series of samples Pr(Fe,Ru)AsO and La(Fe,Ru)AsO. According to ref. [39], this behavior, along with the observed behavior of the Hall voltage below 50 K, suggests that the contribution from one of the hole bands may be enhanced in this material at low temperature. However, an alternative explanation is that for low levels of chemical substitution, the effect of disorder dominates over the effect of band structure changes, resulting in a complete suppression of the magnon drag minimum around 50K. In this description, the Seebeck curve of the $x = 0.10$ sample replicates the behavior of disordered undoped parent compounds in the low temperature regime, assuming positive values determined by diffusive multiband Seebeck contributions, as previously discussed in this section. At large $x \geq 0.5$ the negative contribution to S dominates. Hence Ru doping tunes the competition between electron and hole bands in different regimes of temperature and doping.

Gradual suppression of the SDW transition can also be studied by progressively substituting the rare earth as in (La,Sm)FeAsO and (La,Y)FeAsO series, as displayed in the middle panel of Figure 4. The substitution tunes the $T_N$ temperatures between the values of the end members of each series and the features in the Seebeck curves in correspondence of the transitions are shifted along the temperature axis, without major changes in the curve shapes. This is seen clearly by properly rescaling the curves and plotting them versus the reduced temperature $T/T_N$. On the other hand, the Seebeck peak at 50K-70K attributed to magnon drag is not appreciably affected by the rare earth substitutions, indicating that disorder on rare earth site does not provide scattering centers for magnons, oppositely to the case of disorder on the Fe site, for example in the above mentioned LaFe$_{1-x}$Ru$_x$AsO doped series, where the magnon drag suppression seems to be much more effective. This observation is pretty reasonable, given that AFM magnons are associated to the ordering of Fe moments.

Zn doping on the Fe site is more effective in suppressing the SDW transition as compared to Ru doping [40]. Indeed, in LaFe$_{1-x}$Zn$_x$AsO a drastic suppression is observed at low $x \leq 0.1$, as shown in the right panel of Figure 4. This doping value is also the solubility limit, as for larger substitutions phase separation occurs. It can be also noted that the magnon drag minimum around 50K is completely suppressed for doping values as low as $x=0.02$, confirming the dramatic effect of disorder on the magnum drag mechanism. Interestingly, whereas Zn$^{2+}$ ions affect significantly the AFM order, the disorder caused by Zn doping has little effect on the superconducting electron pairing in the corresponding F doped compound LaFe$_{1-x}$Zn$_x$AsO$_{0.9}$F$_{0.1}$.

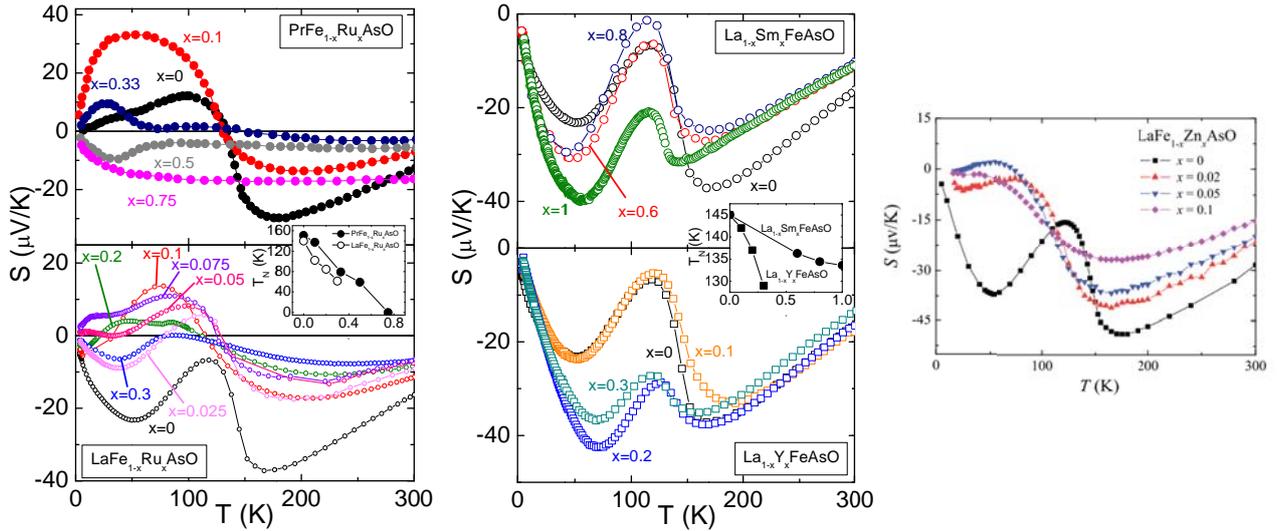

**Figure 4:** Behavior of doped but not superconducting compounds of the 1111 family. Left: Seebeck coefficient of PrFe$_{1-x}$Ru$_x$AsO polycrystals from ref. [39] and LaFe$_{1-x}$Ru$_x$AsO polycrystals prepared as in ref. [41]. In the inset, the trend of $T_N$ extracted from the corresponding features in the transport curves is plotted as a function of Ru content x for both series of samples. Middle: Seebeck coefficient of La$_{1-x}$Sm$_x$FeAsO and La$_{1-x}$Y$_x$FeAsO polycrystals prepared and characterized as in ref. [42]. Right: Seebeck coefficient of LaFe$_{1-x}$Zn$_x$AsO polycrystals from ref. [40] (figure is taken from ref. [40], copyright 2009 by Institute of Physics).

To conclude this section, it is interesting to compare the behavior of 1111 parent compounds with that of akin compounds that do not exhibit the SDW transition. LaFePO is non-magnetic, non-superconducting and does not exhibit any SDW transition. In ref. [43], the difference between LaFeAsO and LaFePO is discussed in terms of the different shape of the metal-pnictide tetrahedron between the P and As cases. Indeed Fe-Fe distance is larger in LaFeAsO than it is in LaFePO, meaning that the former should be closer to localized (magnetic) electron behavior. The larger size of As with respect to P results in a lower compression of tetrahedra in LaFeAsO than in LaFePO, so that Jahn-Teller and Hubbard splittings drive the As system closer to localized, non-metallic behavior, thus favoring the transition to the AFM ordering. The thermopower in LaFePO is small and negative, around -20 µV/K at room temperature, consistently with LaFePO being a n-type metal.

LaNiAsO is the parent compound of iron(nickel)-based layered superconducting compounds [44]. Neither structural transition nor AFM ordering associated with Fe(Ni) ions occurs in LaNiAsO, and no nesting characterizes its Fermi surface. Both normal state properties and superconductivity of the NiAs-based systems are likely of conventional type. Also in the parent compound LaNiAsO, S is very small, only few µV/K, as typical of metals.

**4.2 Seebeck effect of 122 parent compounds**

In the case of 122 compounds, the availability of mm size single crystals allows better-quality and finer thermoelectric measurements, with the further possibility of investigating anisotropy and nematicity. Very similarly to the case of their 1111 counterparts, the 122 parent compounds undergo crystallographic and magnetic transitions at temperatures between 140 K and 200K. In Figure 5, we present Seebeck curves of different 122 parent compounds collected from literature, namely BaFe$_2$As$_2$ [45,46,47], CaFe$_2$As$_2$ [48,49] and EuFe$_2$As$_2$ [50,51].

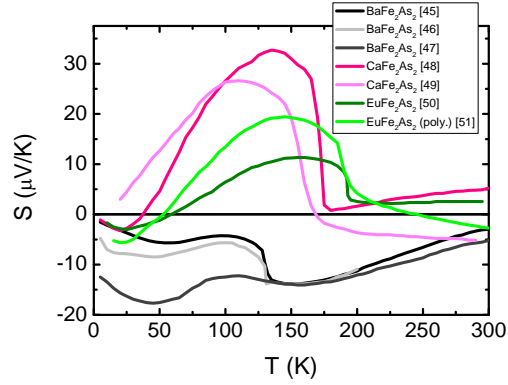

**Figure 5:** Seebeck coefficient curves of $AFe_2As_2$ (A=Ba, Ca, Eu) samples taken from ref. [45,46,47,48,49,50,51]. All the samples are single crystals except for the polycrystal of ref. [51].

It can be clearly seen that, similarly to the case of 1111 parent compounds, the Seebeck effect curves of samples with the same composition may depart significantly below $T_{SDW}$, especially the $BaFe_2As_2$ single crystals of ref. [45,46,47]. However, differently from 1111 parent compounds, an appreciable spread is also observed in the temperature range above $T_{SDW}$. This suggests that 122 parent compounds are very close to compensation and the multiband transport properties are very sensitive to the contribution of each band. In all cases, S curves undergo an abrupt jump at the magnetic/structural transition toward more positive S values and exhibit non monotonic behavior in the low temperature SDW state, where they are characterized by a broad maximum below the magnetic/structural transition at temperatures between 100K and 150K and a minimum between 20K and 50K, before eventually vanishing at the lowest temperatures. In the high temperature regime above the magnetic/structural transition the S curves are featureless and very small in value. Differently from 1111 parent compounds, there is some systematic variability depending on the alkaline earth metal, either Ba, Ca or Eu, both above and below $T_{SDW}$. The Seebeck of $BaFe_2As_2$ is negative in the whole temperature range and it attains a maximum absolute value ~15μV/K around 150 K, just above the transition [45,46,47]. Instead, for the other parent compounds $CaFe_2As_2$ and $EuFe_2As_2$, the Seebeck curves become positive in the low temperature regime, reaching a value of few tens μV/K at the broad maximum just below the transition [48,49,50,51]. It appears that the smaller is the ionic radius of the alkaline earth metal, the larger is the hole contribution to transport with respect to the electron contribution.

The dramatic upturn of S at the transition must be related to the structural transition, not to the magnetic transition, as demonstrated under an applied pressure that splits the two transitions [45]. The S upturn must be attributed to a steep increase of the hole contribution, related to the reconfiguration of electronic structure and appearance of a hole-like band at the structural transition. This scenario is consistent with the one extracted from angle-resolved photoelectron spectroscopy (ARPES) measurements [52].

The non monotonic behavior of S curves below the transition is closely reminiscent of that of 1111 parent compounds. Hence, it is natural to attribute the minimum around 20K-50K to a magnon drag mechanism, superimposed to a multiband diffusive Seebeck contribution that accounts for the broad maximum between 100K and 150K. However, for $EuFe_2As_2$ alternative explanations for the minimum at 20K are suggested, namely it is associated to antiferromagnetic ordering of $Eu^{2+}$ moments [51] or to phonon drag [53].

As the Seebeck effect is a powerful tool to explore effects related to Fermi surface and scattering mechanisms, it turns out to be more sensitive than resistivity to nematicity, the spontaneous symmetry breaking of planar crystalline directions, driven by magnetic or orbital degrees of freedom. Nematicity is observed in transport, magnetic and optical properties of 122 compounds at the boundary between tetragonal and orthorhombic phases, but universal consensus has not yet been achieved about its origin. Nematicity of the Seebeck effect is investigated in detwinned single crystals of the isovalent doped series $EuFe_2(As_{1-x}P_x)_2$ with the aim of clarifying some aspects of this debated issue [54]. Two composition x=0.05 and 0.09 are considered, corresponding to the regime of the phase diagram where structural and magnetic transition occur at $T_s$ and $T_N$, respectively, and no superconductivity develops. In Figure 6 (panels c) and d)), the temperature behaviors of $S_a(T)$ and $S_b(T)$, the Seebeck coefficients along the in-plane a and b axes, are displayed. At high temperature, $S_b(T)$ decreases with decreasing temperature, becoming negative just above $T_s$, at $T_s$ it increases abruptly to positive values and then undergoes a further change of slope at $T_N$. After passing a maximum around 100 K, $S_b(T)$ then decreases and displays a broad negative minimum near 25 K. In comparison corresponding features in $S_a(T)$ are much weaker. More remarkably, $S_a(T)$ and $S_b(T)$ cross at $T_s$, so that the Seebeck anisotropy, defined as $(S_b-S_a)$ and plotted in Figure 6 (panels e) and f)), changes in sign at $T_s$. The authors consider the Mott relationship expressed as a sum of two contributions reflecting the anisotropies of the scattering time and of Fermi surface, respectively. The former is affected by anisotropic scattering due to magnetic fluctuations in the paramagnetic state above $T_s$, while the latter is influenced by orbital polarization and Fermi surface reconstruction below $T_N$. Indeed, below $T_N$, orbital polarization develops as a consequence of the shift of the density of states below the Fermi energy along the b-axis, while in the paramagnetic state above $T_s$ the scattering of carriers is anisotropic near hot spots of the Fermi surface, connecting electron and hole pockets. The sign change of Seebeck anisotropy is thus explained by the competition of two contributions, namely anisotropic fluctuations, dominating at high temperatures ($T>T_s$) and orbital polarization, dominating at low temperatures ($T<T_N$). A similar investigation on superconducting $EuFe_2(As_{1-x}P_x)_2$ with x=0.23 does not evidence any in plane anisotropy [54].

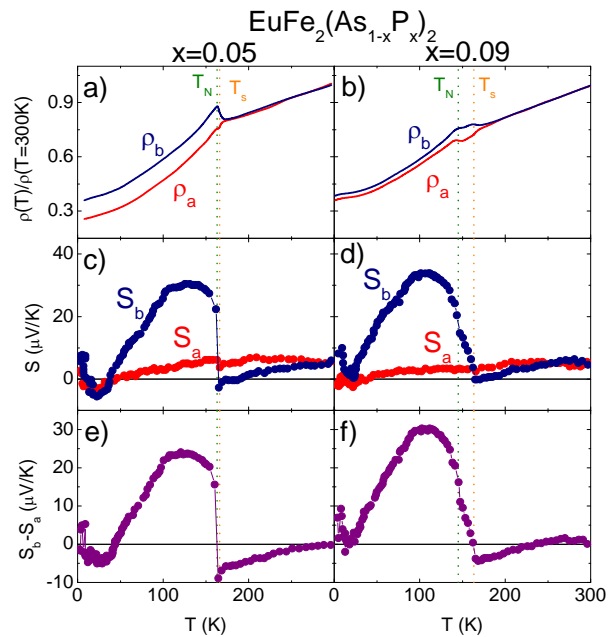

**Figure 6:** Electrical resistivity (a) and (b) and Seebeck coefficient (c) and (d) of $EuFe_2(As_{0.95}P_{0.05})_2$ and $EuFe_2(As_{0.91}P_{0.09})_2$ single crystals along the orthorhombic a and b axes, indicated by red and blue, respectively. In the bottom panels e) and f), temperature dependences of the Seebeck coefficient anisotropy $(S_b-S_a)$ are plotted for the two samples. Dotted vertical lines indicate the structural phase transitions at $T_s$ and the antiferromagnetic transitions at $T_N$, respectively. Data taken from figures 2 and 4 in ref. 54.

The effect of an applied pressure in tuning the phase diagram of the parent compound $BaFe_2As_2$ is extensively investigated in ref. [45] and shown in Figure 7. In many respects, pressure is found to be very similar to electron doping by Co substitution. First of all, an applied pressure P>1.5 GPa splits magnetic and structural transitions, in analogy with Co doping [55]. This allows to associate unambiguously the upturn of S to the structural transition, also visible as a change of regime in the resistivity curves, rather than to the magnetic transition. Secondly, applied pressure shifts the Seebeck curves to more negative values over the whole temperature range and decreases the upturn of S at structural transition. Both these effects indicate a relative increase of the electron contribution. Indeed the upturn of S is related to the hole contribution consequent to appearance of a hole-like band at $E_F$, below the transition. This positive contribution in S vanishes with increasing pressure, signaling a fall of the hole band below the $E_F$.

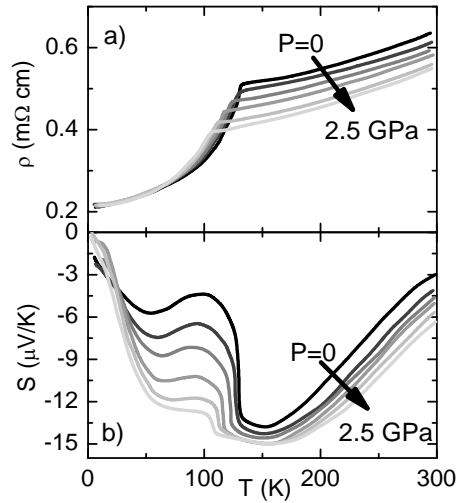

**Figure 7:** (a) Resistivity under pressure and (b) Seebeck coefficient for the parent compound $BaFe_2As_2$. The vertical line at the transition temperature of the ambient pressure curve shows that the sudden increase of the hole contribution to S coincides with the drop in the resistivity curve. Data taken from figure 1 of ref. [45].

### 4.3 Seebeck effect of the 11 $Fe_{1+y}Te$ compound

It is under debate whether $Fe_{1+y}Te$ can be considered the parent compound of the 11 iron-based superconductors Fe(Te,Se). In this respect the actual stoichiometry of Fe is crucial. Excess Fe is thought to dope electrons in the system and be associated with a large magnetic moment ~$2.4\mu_B$ [56]. $Fe_{1+y}Te$ undergoes a structural and magnetic transition at ~65 K from a tetragonal paramagnetic to an orthorhombic AFM state. With decreasing excess Fe and increasing substitution of S or Se in the Te site, the AFM order with wavevector $(\pi,0)$ is gradually suppressed and $(\pi,\pi)$ spin fluctuations associated to the Fermi surface nesting vector increasingly replace $(\pi,0)$ spin fluctuations associated to local magnetic moments, thus complying with the mostly accredited picture of unconventional superconductivity in iron-based families. Above the structural/magnetic transition, the transport behavior of $Fe_{1+y}Te$ exhibits weakly temperature dependent resistivity, at the transition the resistivity drops sharply and below the transition a metallic state sets on [36]. The first Seebeck effect curve of $Fe_{1+y}Te$ ever reported, measured on a single crystal [57], is shown in Figure 8. Also shown are a very similar curve measured in another single crystal [58] and a curve measured on a polycrystalline sample [36]. The difference between the single crystals and the polycrystal Seebeck curves could be related to the different excess Fe, which should dope electrons in the system and shift the Seebeck curve toward negative values. Actually, the measured Fe stoichiometry in the three samples does not support this view, but the different techniques used in each case to evaluate the Fe content introduces a large uncertainty in the direct comparison among the samples. As compared to the Seebeck curves of other Fe-based parent compounds, in this case S assumes very small values in the whole temperature range. Another peculiarity is the distinctive flat temperature dependence of S above the transition, with vanishingly small value S≈-1 µV/K, in the case of the single crystals. At the magnetic/structural transition, S undergoes an abrupt step-like change, it reaches a minimum value S≈-11-12 µV/K around 50K and eventually tends to vanish as the temperature tends to zero. The negative sign of S in the whole temperature range is at odds with the sign of the Hall resistance $R_H$, which is positive above the transition [36]. This situation is a clear evidence of multiband transport, as it typically occurs in case of simultaneous presence at the Fermi level of an electron band with smaller density n and higher mobility µ and an hole band higher n and smaller µ. Indeed, roughly speaking, holes (h) and electron (e) bands contributions are weighed differently in the expression for $R_H$ [i], which assumes the sign of $(\mu_h^2 n_h - \mu_e^2 n_e)$, and in the expression for S [ii], which assumes the sign of $(\mu_h n_h S_h - \mu_e n_e S_e)$. Moreover, the small magnitude of S, especially in the high temperature regime, indicates a high degree of compensation between hole and electron bands (S would vanish in the ideal case $n_e=n_h$), confirmed also by very small values of the Hall resistance [36]. The flat behavior above the transition is well described by the Heikes law eq. (8) and indicates that $Fe_{1+y}Te$ is narrow-banded with strong on-site Coulomb repulsion, namely $U>>k_BT$ in the flat S regime. As for the explanation of the S jump at the magnetic/structural transition, similar possibilities can be considered as in the case of the other iron-based parents compounds, the most likely is the reconstruction of the Fermi surface, but a change in the scattering mechanism may also play a role.

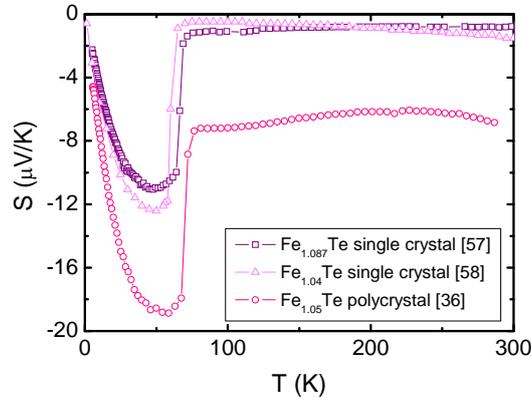

**Figure 8:** Seebeck coefficient for a $Fe_{1.087}Te$ single crystals, taken from ref. [57], $Fe_{1.04}Te$ single crystals, taken from ref. [58] and $Fe_{1.05}Te$ polycrystals taken from ref. [36].

Below the structural/magnetic transition, the non monotonic behavior of S needs a more articulated analysis. The minimum at 50K could be interpreted as originating from a drag mechanism, either phonon or magnon drag. Regarding the magnon drag scenario, it can be ruled out because (i) 50K is too high a peak temperature for magnon drag, given that the magnetic transition temperature is only slightly larger ~65K, and because (ii) no significant magnetic field dependence of S is detected in this temperature regime [57,58]. The missing signatures of magnon-drag suggest that the spin fluctuations related to AFM ordering in $Fe_{1+y}Te$ do not couple significantly with charge carriers. This scenario matches with the experimental [59,60,61] and theoretical [62] findings that in $Fe_{1+y}Te$ the Fe moments align according to a magnetic wave vector (π,0), in contrast with the AFM order along the nesting wave vector (π,π) of 1111 and 122 parent compounds. While the (π,π) spin fluctuations couple with carriers [63,64], (π,0) spin fluctuations are not expected to, because they do not match any nesting wave vector [65,66]. With respect to the possibility of a phonon drag contribution, the peak

---

[i] The two band expression for $R_H$ is $R_H = \frac{1}{q}\frac{(\mu_h^2 n_h - \mu_e^2 n_e)}{(\mu_h n_h - \mu_e n_e)^2}$, where the subscripts *h* and *e* indicate hole and electron bands respectively and *q* is the electron charge.

[ii] The two band expression for S is $S = \frac{\sigma_h S_h + \sigma_e S_e}{\sigma_h + \sigma_e}$, as in eq. (7), where $\sigma=qn\mu$ are the conductivities of the two bands. As S is inversely proportional to n in the diffusive regime, it turns out that each band contribution is mostly weighed by the respective mobility *µ*.

temperature ~50K is compatible with a Debye temperature ≈200K in this compound [67,68]. However, the broadness of the peak rather suggests a diffusive multiband mechanisms, resulting from temperature dependent relative contributions of electron and hole bands as well as from vanishing entropy as the temperature approaches zero.

First-principles density functional theory (DFT) calculations of structural, electronic, magnetic and transport properties of FeTe, including Seebeck coefficient, are presented in ref. [69]. They reproduce the signature of the structural/magnetic transition in FeTe, namely discontinuity and sign change of the Seebeck coefficient, provided that antiferromagnetic correlations are taken into account and experimental values of lattice constants are considered in DFT calculations. On the other hand, the temperature dependences of S and the other transport properties cannot be fully reproduced, suggesting that it may be necessary to go beyond the constant relaxation time approximation and take into account correlation effects.

## 5. Seebeck effect of superconducting iron pnictides and chalcogenides

In general the Seebeck effect is a probe of electronic properties and in the diffusive regime it is not significantly affected by extrinsic mechanisms (see eq. 6). For this reason it is a powerful tool to extract quantitative information on the effect of doping in single band systems, for example it has been used to investigate the charge transfer mechanism in high-$T_c$ cuprates, where a universal relationship between the room temperature Seebeck value and the amount of charge per $CuO_2$ plane has been established [70,71]. This approach is not applicable to the case of multiband systems, where the Seebeck coefficient also depends on the conductivity of each band (see eq. 7). For this reason, in iron-based superconductors only qualitative analyses of Seebeck curves are possible.

## 5.1 Seebeck effect of 1111 superconductors

The most studied doping of the 1111 family superconductors is F substitution on the O site, which is also the original substitution that lead to the discovery of superconductivity in iron-based compounds [2]. F dopes electrons in the system, yielding a dome-shaped phase diagram characterized by an optimal $T_c$ for 15% substitution. From transport properties [36], it is generally observed that F doping suppresses the structural/magnetic transition, lowers the resistivity and yields a metallic state that turns into superconducting state at $T_c$ up to ~55K. Hall effect data show that the carrier density and mobility of the optimally doped samples are rather weakly temperature dependent. The Seebeck curves of undoped and F-doped SmFeAsO samples, taken from ref. [36] and [72], are shown in Figure 9. S is negative over the entire temperature range, in agreement with the Hall coefficient data. A non-monotonic behavior of the Seebeck effect with doping is observed, namely |S| increases in absolute value from x=0 to 0.075 and then decreases for larger x=0.15 and x=0.2. This non-monotonic behavior, observed also in doped LaFeAsO [32,73] and in superconductors of the 122 family (see section 5.2), can be understood considering the almost compensated nature of these compounds and the multiband character described by eq. (7), as explained in the following. The x=0 sample shows smaller S values since it is the most compensated, so that electron and hole contributions almost cancel out. With F doping, the electron contribution prevails more and more, so that |S| increases in magnitude and becomes more negative (x=0.075 sample). Eventually, when the doping is large enough, in eq. (7) the hole term can be neglected at all and S can be approximately described by a single band contribution, with |S| decreasing with increasing doping $S \propto 1/n$ (x=0.15 sample), according to eq. (6).

The S(T) behavior of other F doped $RE$FeAsO compounds is qualitatively similar to that displayed in Figure 9 for the F doped SmFeAsO compound, namely S is negative and increases in magnitude with decreasing temperature, it reaches a broad minimum at a temperature between 70 and 120K and finally decreases in magnitude with decreasing temperature, down to the sharp drop to S=0 at the superconducting transition (see for example data measured on $NdFeAsO_{0.9}F_{0.1}$ [72] and $LaFeAsO_{1-x}F_x$, where x=0.05 [32], 0.1 [32], 0.11 [74], 0.15 [72]). More generally, this S(T) shape actually describes also different types of n-type doping of 1111 compounds, as clearly seen in Figure 10a), reporting S data of polycrystalline TbFeAsO$_{0.85}$ (T$_c$=42.5K) [75], SmFeAsO$_{0.85}$ (T$_c$= 53K) [76], La$_{0.8}$Th$_{0.2}$FeAsO (T$_c$=30.3K) [77], Tb$_{1-x}$Th$_x$FeAsO (T$_c$=50K) [78], LaO$_{0.8}$F$_{0.2}$FeAs$_{0.95}$Sb$_{0.05}$ (T$_c$=30.1K) [79], LaFe$_{0.89}$Co$_{0.11}$AsO (T$_c$=14.3K) [80], LaFe$_{0.9}$Co$_{0.1}$AsO (T$_c$=13K) [81], SmFe$_{0.9}$Co$_{0.1}$AsO (T$_c$=17.2K) [81], CeFe$_{0.9}$Co$_{0.1}$AsO (T$_c$=12.5K) [82], SmFe$_{1-x}$Ni$_x$AsO, (T$_c$=10K for x=0.06) [83], PrFe$_{0.9}$Co$_{0.1}$AsO (T$_c$=16K) [84], LaFe$_{1-x}$Zn$_x$AsO$_{0.9}$F$_{0.1}$ (T$_c$~25K for x=0.02 and 0.05) [40], Sm$_{0.9}$Th$_{0.1}$FeAsO$_{0.85}$F$_{0.15}$ (T$_c$~55K) [85]. These S(T) curve shapes are typical of low charge density metals. With increasing doping of 1111 compounds, in the overdoped regime, the broad minimum shifts to higher temperatures, eventually above 300K, so that the S(T) curves increase monotonically and almost linearly with increasing temperature up to 300K [82,83,84], approaching the behavior of conventional metals, as shown in Figure 10b). A similar trend of overdoped 122 compounds is described in section 5.2 and shown in the right panel of Figure 12. Discriminating between possible different mechanisms determining the shape of S(T) curves of optimally doped 1111 samples shown in Figure 10a) is not easy. It is possible that the multiband diffusive Seebeck alone, without any drag contributions, could account for the non monotonic S(T) behavior, due to the competition between dominant electronlike bands and the expected proximity of holelike bands near the Fermi energy [74]. Drag mechanisms are expected to be severely suppressed by doping, as doping introduces disorder in the system. However some authors suggest that the broad minimum is due to the phonon drag mechanism [75,76]. In ref. [76], it is suggested that the phonon drag Seebeck contribution may be enhanced by a resonant mechanism related to the Fermi surface nesting in the parent compound, namely it could involve mainly those phonons having wavevector equal to the nesting vector. This picture could account for the suppression of |S| peak and of T$_c$ with increasing pressure [76]. Indeed, with increasing pressure and consequent shifting away from the nesting condition, the resonant phonon drag Seebeck contribution would accordingly be suppressed. Alternatively or besides phonon drag, also magnon drag may play a role. Indeed, in principle, if a strong coupling between charge carriers and spin fluctuations exists, a sizeable magnon drag contribution to the Seebeck effect may be expected, as observed in the parent compounds (see section 4.1). Unfortunately, no Seebeck measurements in applied magnetic field are available in literature to confirm this hypothesis. However, the above mentioned effect of applied external pressure investigated in ref. [76], could be most plausibly explained in terms of a resonant mechanism involving spin fluctuations rather than phonons, thus recovering the magnon drag scenario demonstrated for parent compounds.

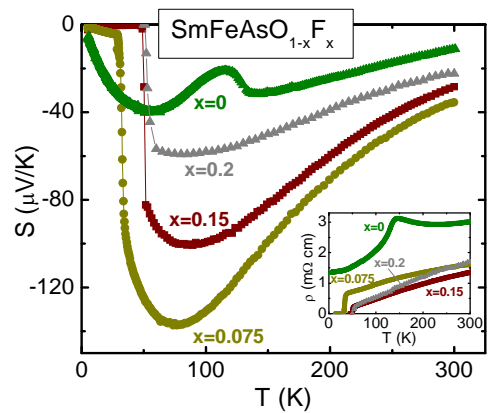

**Figure 9:** Seebeck curves of SmFeAsO$_{1-x}$F$_x$ samples. Data for x=0, 0.075, 0.15 are taken from ref. [36], data for x=0.2 are taken from ref. [72]. Inset: resistivity curves of the same samples.

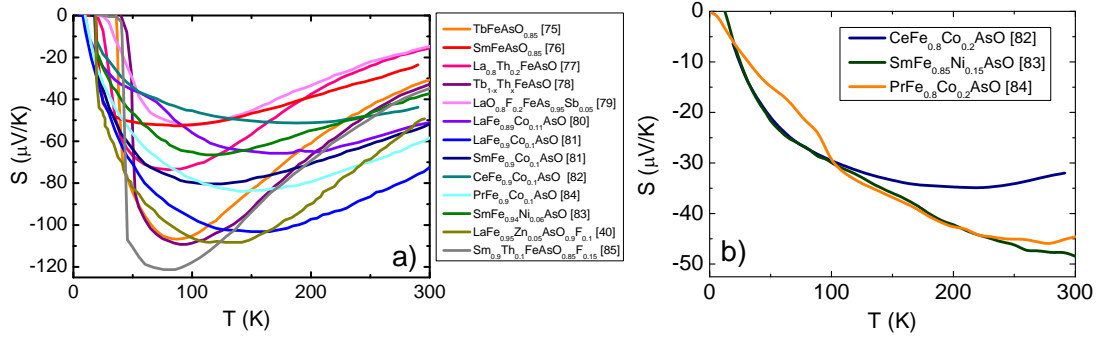

**Figure 10:** a) Seebeck curves of doped 1111 polycrystals at, or close to, optimal doping: TbFeAsO$_{0.85}$ [75], SmFeAsO$_{0.85}$ [76], La$_{0.8}$Th$_{0.2}$FeAsO [77], Tb$_{1-x}$Th$_x$FeAsO [78], LaO$_{0.8}$F$_{0.2}$FeAs$_{0.95}$Sb$_{0.05}$ [79], LaFe$_{0.89}$Co$_{0.11}$AsO [80], LaFe$_{0.9}$Co$_{0.1}$AsO [81], SmFe$_{0.9}$Co$_{0.1}$AsO [81], CeFe$_{0.9}$Co$_{0.1}$AsO [82], PrFe$_{0.9}$Co$_{0.1}$AsO [84], SmFe$_{0.94}$Ni$_{0.06}$AsO [83], LaFe$_{0.95}$Zn$_{0.05}$AsO$_{0.9}$F$_{0.1}$ [40], Sm$_{0.9}$Th$_{0.1}$FeAsO$_{0.85}$F$_{0.15}$ [85]. b) Seebeck curves of doped 1111 polycrystals in the overdoped regime: CeFe$_{0.8}$Co$_{0.2}$AsO [82], SmFe$_{0.85}$Ni$_{0.15}$AsO [83], PrFe$_{0.8}$Co$_{0.2}$AsO [84].

Peculiar features of the Seebeck behavior that emerge in both Zn doped (LaFe$_{1-x}$Zn$_x$AsO$_{0.9}$F$_{0.1}$) [40], Ni doped (SmFe$_{1-x}$Ni$_x$AsO) [83] and Co doped (CeFe$_{0.9}$Co$_{0.1}$AsO, LaFe$_{1-x}$Co$_x$AsO, SmFe$_{1-x}$Co$_x$AsO, NdFe$_{1-x}$Co$_x$AsO$_{0.89}$F$_{0.11}$) [81,82,86] superconducting samples alike, are (i) the close correlation between the absolute value of thermopower and $T_c$, namely, the higher $T_c$ corresponds to larger normal state $|S|$, and (ii) the magnitude of $|S|$ at high temperature that increases with doping, at low doping. As explained above, the latter behavior of substituted 1111 superconductors is accounted for in terms of the multiband character and indicates increasingly dominant electron contribution and departure from the almost compensated condition of the parent compounds with increasing substitution. Indeed, as also confirmed by theoretical calculation, cobalt and nickel substitutions basically shift up the chemical potential, with only minor changes in the density-of-states [87], thus determining increasingly negative S. Electron doping also explain the behavior of $T_c$, indeed the optimally doping level is about 0.06 in the Ni-doped SmFeAsO, nearly half as that of Co-doped SmFeAsO, in fact Ni dopant induces two extra itinerant electrons while each Co dopant only induces one extra itinerant electron. However, the close correlation between the absolute value of thermopower and $T_c$ must involve further mechanisms beside the mere counting of electrons in the conduction band, as it is maintained throughout the full dome along the doping level axis. After subtraction of the diffusive contribution to S, assumed to be due only to the electron bands in the doped samples, the remaining contribution to S, non-vanishing only in the window of the phase diagram where superconductivity occurs, exhibits a clear dome-like behavior as a function of the doping level, mimicking the $T_c$ dome (see Figure 11) [81,82,83]. This seems to be an almost general feature of doped 1111 substituted on the Fe site. The origin of this anomalous contribution to S has not been clarified, but magnetic fluctuations and electron correlations are the most likely candidates [81,82].

In this framework, however, Ru and Mn substitutions on the Fe sites yield different behaviors of the Seebeck curves, which do not follow the above described trend, namely neither the correlation between normal state $|S|$ and $T_c$ nor the increase of high temperature $|S|$ with doping are observed [86,88]. In NdFe$_{1-x}$Ru$_x$AsO$_{0.89}$F$_{0.11}$, the temperature at which the magnitude of S has a maximum remains almost unchanged with increasing Ru doping up to large doping levels x≤0.15 [86], oppositely to the case of Ni and Co doping, where the $|S|$ maximum shifts to higher temperatures with increasing doping [81,82,83,88]. The main difference between Ru substitution and the other mentioned substitutions is that Ru does not donate extra carriers, at least at small doping levels, so that the Fermi surface is not drastically changed. The main effect of Ru substitution is the widening of the *d* bandwidth. Consistently, the magnitude of the normal state Seebeck is lowered by Ru doping and eventually, at large doping x≥0.3, S becomes linearly dependent on the temperature, as expected from a typical metal [86].

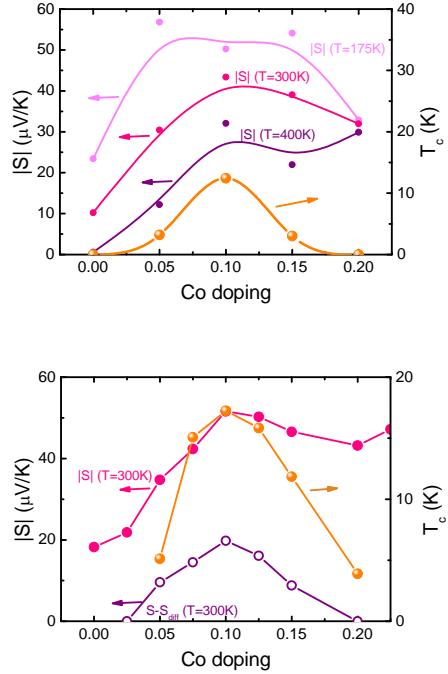

**Figure 11:** Top panel: Co doping content dependence of thermopower at three different fixed temperatures and Co doping content dependence of the critical temperature for CeFe$_{0.9}$Co$_{0.1}$AsO. Data are taken from ref. [82]. Bottom panel: doping dependence of room-temperature thermopower and critical temperature for SmFe$_{1-x}$Co$_x$AsO. The open symbol data are obtained by subtracting the diffusive term form the measured S data. Data and analysis are taken from ref. [81].

Tuning of multiband transport is demonstrated in co-doped samples, where hole and electron doping are introduced simultaneously. In LaFe$_{1-x}$Mn$_x$AsO$_{0.89}$F$_{0.11}$ samples, Mn introduces holes in the system and compensates the electrons induced by F doping, which accounts for the decrease of the magnitude of S, gradually approaching positive values, with increasing x. Although the magnitude of the normal state Seebeck is lowered by Mn doping as in the case of Ru doping, the linear S(T) regime typical of metals is never reached with Mn doping. Indeed, the system become localized at low doping levels (x=0.005 in ref. [88] and 0.08 in ref. [89]) and superconductivity disappears [88,89]. These effects are similar to those observed for Mn doping of 122 samples, described in section 5.2. Another example of co-doping is studied in Pr$_{0.8}$Sr$_{0.2}$Fe$_{1-x}$Co$_x$AsO, where simultaneous electron and hole doping are obtained by Co and Sr substitutions on the Fe and Pr sites of the parent compound PrFeAsO, respectively [84]. Thermopower and Hall data indicate that the hole-type charge carrier dominates in the low Co doping region (x≤0.075) and the system becomes electron type in the high Co doping region (x≥0.075). In particular, the thermopower changes sign from positive to negative with increasing x, evidencing an obvious compensation effect between the electron-type and hole-type dopants. The behavior of $T_c$, which first decreases with increasing Co content to a lowest $T_c$ of 3.5 K at x=0.075 and then increases to a maximum of ~16 K at x=0.15, demonstrates that charge doping plays a key role in superconductivity mechanisms.

Besides doping, applied pressure is used to explore the relationship between superconductivity and structural and electronic properties. The applied pressure can be either external [76], whose effect has is described above in this section, or internal. Internal or chemical pressure is applied by isovalent substitution of P on the As site. A series of SmFeAs$_{1-x}$P$_x$O$_{0.88}$F$_{0.12}$ polycrystals is studied in ref. [90]. P substitution is indeed found to induce anisotropic chemical pressure, mainly shrinking the out-of-plane cell parameter. $T_c$ is suppressed from T=51K for x=0 to $T_c$=20 for x=0.20 and eventually to $T_c$=0 for x=0.30. The Seebeck effect in the normal state is very weakly affected by P substitution in samples with x up to x=0.15 and the broad S(T) minimum is not appreciably shifted in temperature. These results, combined with the analysis of resistivity and Hall effect data, indicate that even upon

P substitution, normal state multiband transport is still dominated by the electron band, consistently with the isovalent nature of the substitution. The suppression of $T_c$ is attributed to structural distortions and disorder in the FeAs planes.

Finally, it is worth considering the magnitude of S generally found in these superconducting 1111 compounds. At optimal doping, the maximum in |S| is even larger than 100 µV/K (see Figure 10a). Such a large value of |S| is quite unusual among superconducting materials. Given that a rough estimate of the conventional diffusive contribution to S according to the Mott expression eq. (5) could barely account for similar values up to ~110 µV/K, the issue of such thermopower enhancement is still open. Strong electron correlation, magnetic fluctuations, peculiar electronic structure could play a role in determining the large magnitude of S. In particular, when the thermal energy $k_B T$ is much lower than the on-site Coulomb repulsion, the contribution of spin entropy to S can be approximately expressed as $S \approx -k_B ln(2)/q \approx 60 \mu V/K$, where the ln(2) term comes from the spin fluctuations of $Fe^{+2}$ ions [75].

It is also pointed out that such large Seebeck coefficient, combined with a rather low resistivity ρ, especially in single crystals and dense polycrystals, yields interesting values of thermoelectric power factor $S^2/\rho$ for thermoelectric applications [72,73].

**5.2 Seebeck effect of 122 superconductors – iron pnictides**

The 122 family has been the most widely investigated among iron-based superconductors in terms of various properties, including thermopower. In this family, superconductivity is induced in the parent compounds upon either electron or hole doping. Also thanks to the availability of sizeable single crystals, the most studied type of electron doping is $BaFe_{2-x}Co_xAs_2$, with maximum $T_c \approx 24K$ for $x \approx 0.11-0.12$, while the most studied type of hole doping is $Ba_{1-x}K_xFe_2As_2$, with maximum $T_c \approx 38K$ for $x \approx 0.32$. However, a large number of other substitutions were explored in the whole doping range and in many cases superconductivity was observed in a doping window around an optimal value. Due to the variety of the observed behavior, different substitutions will be analyzed separately.

In Figure 12, thermopower curves of Co doped $BaFeAs_2$ single crystals in the whole doping range are displayed. Data for the low doping regime $0 \leq x \leq 0.05$ (left panel of Figure 12) are taken from ref. [91]. The S(T) curves evolve gradually to negative values, exhibiting a systematic increase of the magnitude of S with increasing x. For x=0.05 the maximum |S| value is 5 times larger than the maximum value of |S| for x=0. The feature associated to the structural transition, around 140 K in the parent compound x=0, is gradually shifted to lower temperatures and becomes less pronounced, disappearing completely around x=0.034. At this doping level $x \geq 0.034$ the superconductivity jump at $T_c$ appears. The superconducting transition shifts to higher $T_c$ with increasing x, in the so called underdoped superconducting regime. These underdoped S(T) exhibit a broad minimum around 100-110K, which is very similar to the shape of S(T) curves measured in electron doped 1111 superconductors (see Figure 9 and Figure 10a). At higher doping, the trend of increasing |S| with increasing x is inverted, as shown in the middle panel of Figure 12, where data in the intermediate doping regime $0.043 \leq x \leq 0.2$ taken from ref. [46] are reported. This non-monotonic behavior of |S| as a function of doping has already been discussed for the 1111 superconductors in terms of departure from the almost compensated condition of the parent compound at low doping and electron dominated transport with increasing electron concentration at high doping. This picture is consistent also with Hall effect data [92]. In parallel with the decrease of |S| with increasing x, the broad minimum of S is shifted to higher temperatures. This trend continues in the overdoped regime, seen in the right panel of Figure 12, where data for the high doping regime $0.13 \leq x \leq 0.42$ taken from ref. [93] are reported. The broad minimum of S eventually disappears out of the measuring temperature range for x=0.24. At the highest doping x=0.42, S(T) is monotonic with almost linear temperature dependence, as expected for diffusive S of a conventional single band conductor. At this doping value x=0.42, the superconducting transition is fully suppressed.

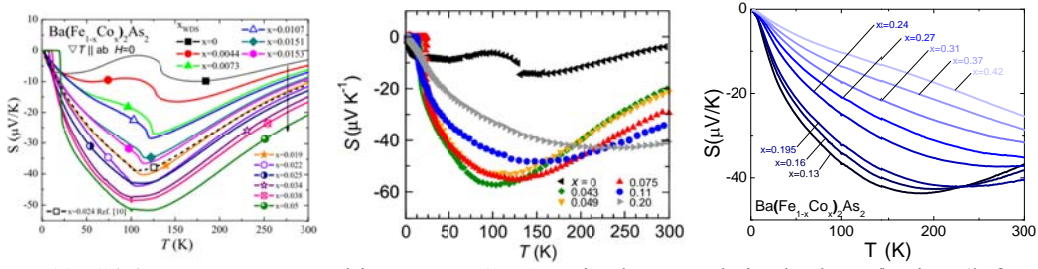

**Figure 12:** S(T) curves measured in BaFe$_{2-x}$Co$_x$As$_2$ single crystals in the low doping (left panel), intermediate doping (middle panel) and high doping (right panel) regimes. Figure in the left panel is taken from ref. [91] (copyright 2013 by Taylor & Francis Ltd., www.tandfonline.com), data in the middle panel are taken from ref. [46], data in the right panel are taken from figure 5 in ref. [93].

The above described phenomenology is consistently observed throughout the available literature reporting Seebeck effect of BaFe$_{2-x}$Co$_x$As$_2$ single crystals [45,46,47,91,92,93].
The shape and magnitude of S for different doping for Co doped BaFe$_2$As$_2$ is well reproduced by a two band model, where transport integrals are solved [94]. This model accounts for |S| around 50-90 microV/K and for the broad maximum in S(T) near 100 K, which is typical of iron-based 1111, 122 and 111 superconducting families. This demonstrates that the key features to account for S are the low carrier concentration semimetal character, the moderately high density of states originated primarily from Fe $d$ bands, the compensating electron and hole Fermi surfaces and very small Fermi energy $E_F/k_B$ =300 K of the undoped parent compound. However, as noted in section 5.1, the even larger S magnitude of 1111 superconductors may be enhanced by contributions from spin entropy and electron correlation.
Seebeck effects measurements were used to investigate the existence of a quantum critical point (QCP) in the region of coexistence between magnetism and superconductivity in the phase diagram of 122 iron-based superconductors. Critical spin fluctuations occurring in the vicinity of the QCP likely play a major role in the superconducting pairing mechanism. In ref. [95], thermopower is measured in a series of BaFe$_{2-x}$Co$_x$As$_2$ samples with closely spaced doping values and a phase diagram is built from the analysis of S/T, allowing the identification of enhanced scattering by critical spin fluctuations close to the antiferromagnetic QCP. Indeed, enhanced spin fluctuations lead to an increase of electronic entropy and, consequently, to an increase of diffusive thermopower, which can be described by the law:

$$S/T \propto (1/q)\left(\frac{g_0^2 N'(0)}{E_F \omega_S N(0)}\right) ln(\omega_S/\delta) \qquad (17)$$

where N(0) is the density of states at the Fermi energy $E_F$, $g_0^2$ is the bare coupling between the electrons and spin fluctuations, $\omega_S$ is the energy of the spin fluctuations and $\delta = \Gamma(p-p_c)+T$ is the mass of the spin fluctuations, which measures the deviation from the QCP, tuned by temperature or by any other parameter p (doping, pressure or magnetic field). The dependence S/T $\propto$ ln(1/T) is thus a signature of the quantum critical behavior, opposed to the Fermi liquid behavior S/T≈constant described by Mott equations. According to S/T curves reported in Figure 13 (left side panels), as a function of the doping x, the authors of ref. [95] divide the phase diagram into three regions, separated by three Lifshitz transitions at x≈0.025, 0.11 and 0.2. At low doping x<0.025, S is positive and undergoes an abrupt change at the structural transition temperature, which is gradually lowered by doping. As x≈0.025 is approached, the quantum critical behavior S/T $\propto$ ln(1/T) is observed in a limited temperature range (30-100 K). At intermediate doping 0.034<x<0.114, S/T increases linearly on the ln(T) scale with lowering temperature. With an increase of x, the slope of S/T logarithmic temperature dependence increases up to x=0.05 and then decreases. At higher doping the slope of S/T decreases further and above x≈0.2 low temperature S/T saturates, as the system makes a crossover from a quantum critical non Fermi liquid to the Fermi-liquid-like (S/T≈constant) state. In summary, the slope of S/T logarithmic temperature dependence attains its highest value at the QCP x=0.05 (see Figure 13 upper right panel) and the x dependence of S/T comes from the change in the spin fluctuation mass δ and the Fermi energy in the expression for S. This picture is supported by the observation of linear temperature dependence of resistivity at doping levels close to the QCP x=0.05. In order to evidence the connection between superconductivity and the observed quantum criticality, the authors show the correlation between the x dependences of three quantities, namely $T_c$, S/T and the specific heat jump at $T_c$ (see Figure 13 lower right panel). This correlation emphasizes the relevance of this analysis and supports the picture of superconductivity mediated by spin fluctuations.

Evidence of critical spin fluctuations from logarithmic temperature dependence of S/T is also obtained on hole doped $(Sr_{1-x}K_x)Fe_2As_2$ [96] and $K_xEu_{1-x}Fe_2As_2$ [53] systems, described later on in this section.

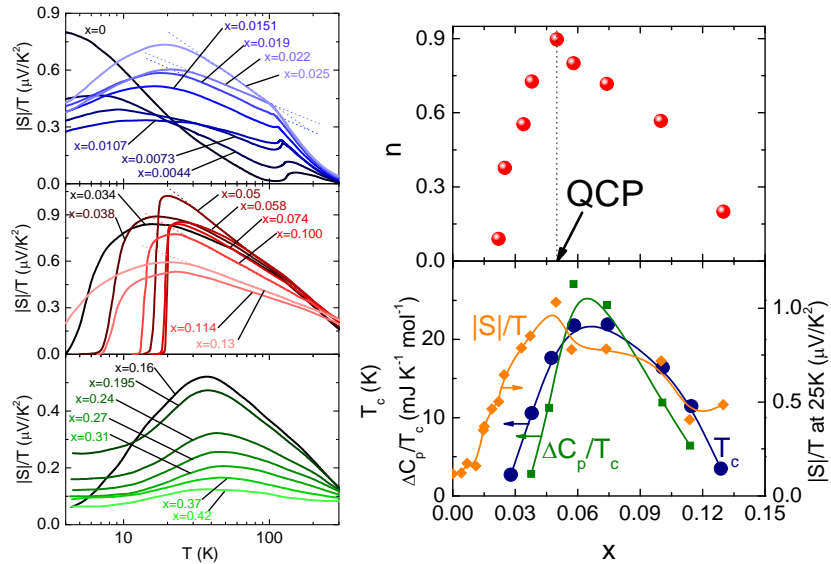

**Figure 13:** Left: S/T versus ln(T) in three Co substitution regimes of the phase diagram of $BaFe_{2-x}Co_xAs_2$. The dotted lines emphasize linearity on a ln(T) scale. Upper right: plot of the slope n of the logarithmic temperature dependence of S/T as a function of Co substitution x, described by eq. (17). Lower right: superconducting transition $T_c$, specific heat jump $C_p/T_c$ and thermopower (S/T) at T=25 K as a function of concentration x. Data and analysis are taken from figure 1 and 2 and 5 of ref. [95].

Measurements of Seebeck effect in series of samples with closely spaced levels of doping allow to identify critical levels of doping at which normal state properties are changed abruptly, likely as a consequence of changes in the Fermi surface and/or band structural properties. ARPES measurements are a direct probe of the existence of such abrupt changes. For example, from ARPES investigations on $Ba(Fe_{1-x}Co_x)_2As_2$ crystals [52], it was found that a hole pocket appearing below the structural transition in the parent compound is suppressed below the Fermi level for x∼0.03, leading to a Lifshitz transition and onset of superconductivity. The relationship between such changes in normal state properties and occurrence of superconductivity is an important clue to gain insight on the superconductivity mechanisms themselves. Different kinds of doping of the 122 phase are considered for such investigation. In ref. [92], measurements of Seebeck and Hall effect of $BaFe_{2-x}Co_xAs_2$ and $BaFe_{2-x}Cu_xAs_2$ single crystals, the latter showing no evidence of superconductivity, are analyzed. Critical doping values x≈0.022 for Co and x≈0.0085 for Cu are found, which both correspond to e≈0.020 extra electrons, on the basis of valence arguments (e≈x for

Co and $e\approx 3x$ for Cu). Whereas in the case of Co doping the change in the overall form of the S and Hall data takes place as the sample is doped into the region of $e$ values that supports superconductivity, in the case of Cu doping such change occurs at $e$ values that overshoot the superconducting range. In other words, at doping values where the structural/magnetic transitions are suppressed enough, the Cu-doped samples are too overdoped for superconductivity to be established. This is confirmed by the slower rate of structural/magnetic transitions as a function of induced extra electrons in Cu doped samples as compared to Co doped sample and indicates that the above described changes of normal state properties are not a sufficient condition for superconductivity to appear. This picture about the relationship between rate of decrease of the structural/magnetic transition and appearance of superconductivity with increasing doping is consistent also with the experimental results of ref. [93], where a similar parallel investigation on $BaFe_{2-x}Co_xAs_2$ and $BaFe_{2-x}Ru_xAs_2$ single crystals is carried out, and with the results on hole doped $Ba(Fe_{1-x}Mn_x)_2As_2$ [97] described later on. If a similar scenario is also applied to doped 1111 parent compounds, it turns out that it could account for the absence of superconductivity in $PrFe_{1-x}Ru_xAsO$ [39], where features of the structural/magnetic transition are still visible at x as high as 0.5, but it does not explain the behavior of Zn doped 1111 parent compounds $LaFe_{1-x}Zn_xAsO$ [40], where the structural/magnetic transition is completely suppressed already at $x\leq 0.1$ (see Figure 4).

In ref. [93], evidences for Lifshitz transitions are comparatively investigated in $BaFe_{2-x}Co_xAs_2$ and $BaFe_{2-x}Ru_xAs_2$. Seebeck data for $BaFe_{2-x}Ru_xAs_2$ ($0\leq x\leq 0.36$) single crystals do not exceed ~10 µV/K in magnitude and multiple, broad, features and sign changes are observed for many Ru concentrations. For $0.21\leq x\leq 0.36$, zero Seebeck in the superconducting state appears. Features in S(x) at fixed temperature for x=0.07, 0.2 and 0.3 can be identified and possibly related to Lifshitz transitions or other drastic changes in electronic structure, correlations or scattering mechanisms. For comparison, in the same work [93], features in S(x) are found for Co-doped crystals at x=0.02, in agreement with [92], plus additional features at x=0.11 and x=0.22. Also electron doping in $Ba(Fe,Rh)_2As_2$ single crystals show evidences of Lifshitz transitions. In ref. [91], a comparative study of $Ba(Fe,Co)_2As_2$ and $Ba(Fe,Rh)_2As_2$ single crystals in the low doping region (x<0.05 and x<0.171, respectively) is presented. For $0.026\leq x\leq 0.13$ superconductivity is observed in $BaFe_{2-x}Rh_xAs_2$. The two $Ba(Fe,Co)_2As_2$ and $Ba(Fe,Rh)_2As_2$ systems turn out to be virtually identical in terms of resistance, magnetization, thermopower and phase diagrams. The absolute value of thermoelectric power however is slightly smaller for Rh substituted samples compared to those of Co substituted samples, possible as a consequence of differences in the scattering potentials. Changes of slope in S(x) are observed at x=0.02 for Co doping and around 0.1 for Co and Rh doping and are possibly associated to Lifshitz transitions.

The above described analyses of S(x) dependence in different 122 doped compounds are summarized in the plot of Figure 14, left panel. In order to avoid obvious features related to superconductivity and SDW signatures, S(T=150K) values are chosen, because at T=150K neither superconductivity nor SDW are present in the considered x range. For Ru [93], Cu [92] and Co [91,92,93] doped $BaFe_2As_2$, a Lifshitz transition seems to occur at low x (x≈0.02 for Co and Ru, x≈0.01 for Cu), corresponding to onset of superconductivity for Co, but with no corresponding superconductivity for Ru [93] and Cu [92]. The S(x) data points around this low x region are magnified in the inset in the left panel of Figure 14 and the some of the corresponding S(T) curves for x values across the Lifshitz transitions are shown in the right panel of Figure 14. On the other hand, for Rh [91] doped $BaFe_2As_2$, superconductivity is observed for x=0.026, but no evidence of Lifshitz transition is detected at this doping value. These data possibly suggest that such Lifshitz transition favors the occurrence of superconductivity, but it is neither strictly necessary nor sufficient condition. Additionally, a second anomaly in the phase diagram region corresponding to complete suppression of the structural/magnetic transition seems to be common to several FeAs based compounds.

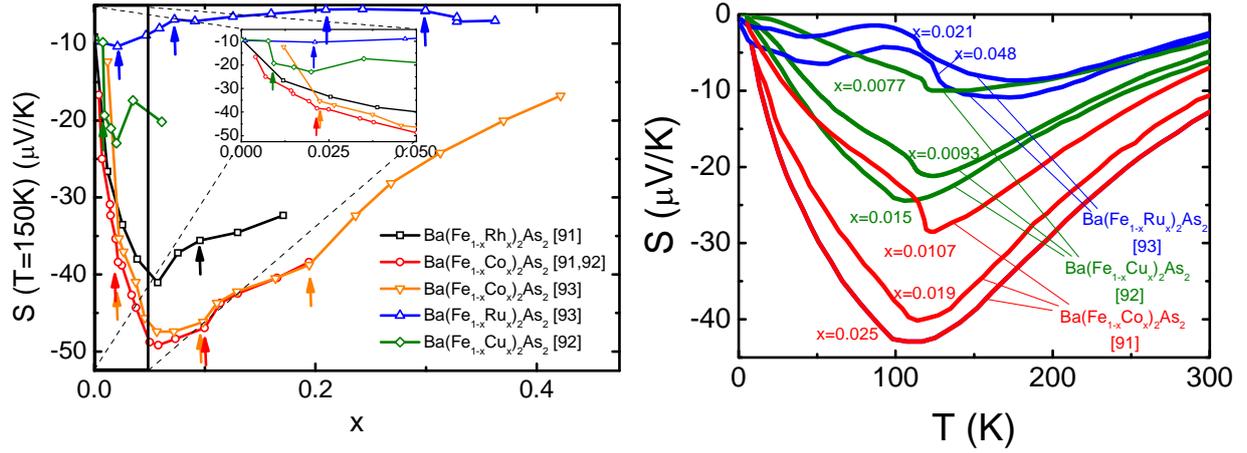

**Figure 14:** Left: S(T=150K) as a function of doping level concentration in BaFe$_{2-x}$Co$_x$As$_2$ [91,92,93], BaFe$_{2-x}$Rh$_x$As$_2$ [91], BaFe$_{2-x}$Cu$_x$As$_2$ [92] and BaFe$_{2-x}$Ru$_x$As$_2$ [93], single crystals. In the inset, the low doping region is magnified. Features possibly related to Lifshitz transitions are indicated by arrows. Right: S(T) curves for x values across the low doping Lifshitz transition from which S(T=150K) data are taken. From the S curves x=0.0077, x=0.0093 and x=0.015 for BaFe$_{2-x}$Cu$_x$As$_2$ [92], x=0.0107, x=0.019, x=0.025 for BaFe$_{2-x}$Co$_x$As$_2$ [91], x=0.021, x=0.048 for BaFe$_{2-x}$Ru$_x$As$_2$ [93], it is clear that abrupt changes of S curve shape and magnitude occur within the narrow doping intervals 0.0077<x<0.0093 for BaFe$_{2-x}$Cu$_x$As$_2$, 0.0107<x<0.019 for BaFe$_{2-x}$Co$_x$As$_2$, 0.021<x<0.048 for BaFe$_{2-x}$Ru$_x$As$_2$, respectively.

From the results presented above, it comes out that Cobalt doping, and more generally any kind of doping in these 122 compounds, cannot be viewed as mere rigid band filling, due to the complex interplay of magnetic, structural and electronic mechanisms. Also a picture only focused on the tuning of relative carrier concentrations of hole and electron bands would be an oversimplification. In ref. [48], the Seebeck, resistivity and Hall effect curves of the Cobalt doped CaFe$_{1.92}$Co$_{0.08}$As$_2$ are compared with the parent compound CaFe$_2$As$_2$ (shown in Figure 5). The authors conclude that above T$_{SDW}$ the main effect of Cobalt doping is not changing carrier concentrations, but rather introducing additional scattering of the hole-like charge carriers, thus favoring electron transport. Below T$_{SDW}$, the doped and undoped curves depart dramatically, the latter undergoing sharp increase to positive values at T$_{SDW}$ and the former having a broad negative minimum around 110K with S≈-20μV/K, similar in shape to the curves of Ba(Fe,Co)$_2$As$_2$ samples.

Hole doping of 122 compound, mostly carried out by K substitution of the alkaline earth metal A in the parent compound AFe$_2$As$_2$, generally yields a gradual lowering of the structural/magnetic transition and a dome-like dependence of the superconducting T$_c$ upon doping in the phase diagram, with the end member still superconducting. In (Sr$_{1-x}$K$_x$)Fe$_2$As$_2$ (0≤x≤1) [98,96], the substitution of K for Sr leads to the onset of superconductivity at x=0.17, an increase of T$_c$ to a maximum T$_c$=37 K near the optimal doping (x=0.42-0.45) and, with further increasing x, a decrease of T$_c$ to 3.8 K at x=1 (KFe$_2$As$_2$). A similar dome-like dependence of T$_c$(x) is found in hole doped (Ba$_{1-x}$K$_x$)Fe$_2$As$_2$ [99,47] and K$_x$Eu$_{1-x}$Fe$_2$As$_2$ [53].

Seebeck curves of K-doped BaFe$_2$As$_2$ samples are shown in Figure 15. In contrast with the negative S of the parent compound (see also Figure 5), S is positive for x values as small as 0.1. S curves exhibit a maximum around 125K [47,99,100], which becomes increasingly broadened and shifts to higher temperatures with increasing doping level, eventually collapsing into a plateau for x≥0.82, accompanied by the appearance of a maximum at ~60 K for pure KFe$_2$As$_2$ [47,99]. The magnitude of S at room temperature, |S(300K)| has a dome-like dependence on the doping level x for K-doped BaFe$_2$As$_2$ [47], at odds with the monotonic |S(300K)| increase of Co-doped BaFe$_2$As$_2$ [47], which is reminiscent of the behavior of high-T$_c$ cuprates. These findings have no straightforward explanation. The authors suggest that spin fluctuations enhance spin entropy and thus contribute to the large S magnitude, but may have a different weight in Co doped and K doped compounds.

Analysis of S(x) dependence at fixed temperatures in a series of crystals with closely spaced doping levels in the overdoped regime indicates a minimum at x~0.55 and a feature at x~0.8-0.9 [99], which the authors associate to Lifshitz transitions corresponding, respectively, to the shift of the electron pockets at the M point above the Fermi level and the transformation of the hole pockets near the M point into "four blades" as observed by experiments of ARPES [101,102] and Hall effect [103] and predicted by theory [104].

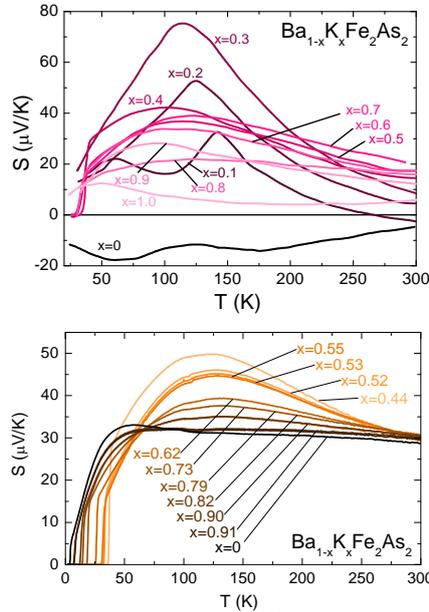

**Figure 15:** S(T) curves of $(Ba_{1-x}K_x)Fe_2As_2$ polycrystals [47] and single crystals having closely spaced doping levels in the overdoped regime [99]. Data in the top panel are taken from figure 4 of ref. [47]. Data in the bottom panel are taken from figure 3 in ref. [99].

As described above about electron doped $BaFe_{2-x}Co_xAs_2$ crystals [95], Seebeck curves provide evidence of critical fluctuations and their possible role in mediating the superconducting pairing. This is also the case of hole doped $(Sr_{1-x}K_x)Fe_2As_2$ samples [98]. In ref. [98], the magnitude of the S maximum $S_{max}$ is traced as a function of doping. Below x=0.3, the SDW order appears and the sharp and distinct peak in $S_{max}(x)$ at x=0.3 indicates a dramatic change of the Fermi surface and the density of states with the onset of the SDW order. The authors also suggest the possible existence of a QCP at this composition, despite the doping value x =0.3 slightly departs from the optimal doping x=0.45.

A comparative analysis of resistivity and Seebeck curves in the same system $(Sr_{1-x}K_x)Fe_2As_2$ shows that at low and high doping the system behaves as a Fermi liquid, but while approaching the critical doping x≈0.4 resistivity becomes linear in T and S/T exhibits logarithmic T dependence (eq. (17)) in a broad temperature range from above superconducting $T_c$=36K to 180K [96]. The results are consistent with the temperature dependencies expected for resistivity and Seebeck curves near a magnetic QCP. Such scenario also applies to $K_xEu_{1-x}Fe_2As_2$ [53], where S(x) at fixed temperature increases rapidly with hole doping and peaks at the critical composition x=0.3, where electronic changes are thought to occur. The electrical resistivity data indicate non Fermi liquid behavior near x=0.5, supported by the observation of a logarithmic divergence of S/T at the same composition. The extended temperature range of critical fluctuation is due to the large energy scale of magnetic fluctuations ($T_{SDW}$=200K) as compared to other systems such as the heavy fermions. It is pointed out in ref. [96] that logarithmic scaling of S/T is expected only for d=2 and z=2 or d=3 and z=3 (d=dimensionality, z= dynamical critical exponent). Which of these cases applies to pnictides is yet to be ascertained.

Hole doping is also explored by substitution on the Fe site in $Ba(Fe_{1-x}Mn_x)_2As_2$ single crystals [97]. Single phase samples are obtained only for x<0.15, while immiscibility occurs for larger x. Differently from electron doping in the Fe sites with various elements such as Co, Ru and Rh, no superconductivity appears with Mn hole doping and the structural and magnetic transitions do not split. The feature in S associated to the structural magnetic transition is gradually lowered by Mn doping. As x increases, there is a shift from purely negative values of S for x<0.033 to increasingly positive values of S at low temperature for x≥0.10. Between x=0.092 and x=0.102 a weak feature of S versus x suggests a Lifshitz transition at this concentration. The authors conclude that Mn substitution is similar to Cr substitution, but different from Co, Ni and Cu. Indeed the suppression of $T_{SDW}$ is much slower with x for Mn and superconductivity does not appear, as observed in the case of electron doped $BaFe_{2-x}Cu_xAs_2$ [92].

It is interesting to compare the effects of doping, chemical pressure and external pressure on the phase diagram and Seebeck response of 122 compounds. In ref. [53], a direct comparison is carried out between hole doped $K_xEu_{1-x}Fe_2As_2$ and isovalent doped $EuFe_2(As_{1-y}P_y)_2$ single crystals. In the latter crystals, P substitution exerts a chemical pressure. In $K_xEu_{1-x}Fe_2As_2$, similarly to the case of other hole doped 122 compounds, the SDW transition is gradually decreased by doping and for x<0.3 superconductivity appears, with maximum $T_c$=34K for x=0.5. S is positive at room temperature. Approaching the SDW transition S(T) exhibits a strong increase leading to a pronounced maximum, peaking around 150 K. At temperatures below 50 K the thermopower crosses over to negative values and exhibits a minimum at 20K. On the other hand, in $EuFe_2(As_{1-y}P_y)_2$ the substitution brings the system from hole dominated to electron dominated transport. Superconductivity is confined to a very narrow regime (0.16<y<0.22). As seen in Figure 16, for y≥0.21 the thermopower is negative at all temperatures, does not have signature of the SDW transition and its magnitude |S| exhibits a broad maximum at 50-150K, which shifts to higher temperatures with increasing y. It is suggested that at y=0.21 a Lifshitz transition occur, in agreement with ARPES experiments [105]. Moreover, logarithmic divergence in the thermoelectric power coefficient S/T suggests that at dopings close to y=0.21 the system is close to a QCP. Accordingly, resistivity temperature dependence $\rho(T)=\rho_0+AT^n$, with n<2, indicates non Fermi liquid behavior [54,53].

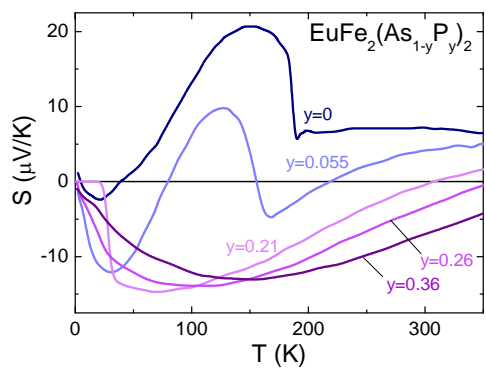

**Figure 16:** S(T) curves of $EuFe_2(As_{1-y}P_y)_2$ single crystals, showing the effect of chemical pressure. Data are taken from figure 7 of ref. [53].

Regarding the effect of applied external pressure on superconducting 122 compounds, experiments carried out on Co doped $BaFe_2As_2$ in applied pressures up to 2.5GPa [45] show that the general trend is similar to that of the undoped $BaFe_2As_2$, in that applied pressure increases the contribution of electrons to transport in both doped and undoped compounds [45]. With increasing pressure, the maximum of |S| is shifted from around 150 K to higher temperatures and slightly decreased in magnitude. The $T_c$ is weakly enhanced by pressure P<1GPa. The overall effect of pressure is thus similar to that of Co doping. Indeed, in the same work [45], by solving standard transport integrals in a two band picture, to fit experimental resistivity and thermopower curves, it is shown that pressure has a similar effect as Co doping on the band structure, namely it pushes the electron band deeper and deeper below the Fermi level, whereas the Fermi level does not change substantially its position with respect to the top of the hole band.

As remarked in the previous section for superconducting 1111 pnictides, it can be said that also for superconducting 122 pnictides, the large absolute values of thermoelectric power and good conductivity, which can be further improved in polycrystals by increasing the density, yield enhanced power factors for thermoelectric applications, namely 1.3 mW m$^{-1}$ K$^{-2}$ in $Ba_{0.7}K_{0.3}Fe_2As_2$ and 1.5 mW m$^{-1}$ K$^{-2}$ in $BaFe_{1.66}Co_{0.34}As_2$ [47]. Indeed, this is a characteristic common to all iron based superconducting pnictides. In the case of the 122 family, the presence of both electron and hole doped compounds suggests possible applications in thermoelectric cooling modules around the liquid-nitrogen temperature range.

**5.3 Seebeck effect of 122 superconductors – iron chalcogenides**

Superconductivity at 32 K has been reported in iron-chalcogenide superconductors $A_xFe_{2-y}Se_2$ (A=K, Rb, and Cs), which share the same crystal structure with iron-pnictides $AFe_2As_2$ (A=Ba, Sr, Ca and K) and can be thought of as FeSe phase intercalated by K, Rb or Cs atoms. Indeed, they are thought to derive from the insulating and magnetic Fe-deficient phase exhibiting Fe vacancy ordering [106], just like the superconducting $FeSe_{1-y}$ phase derives from the $\beta$-$Fe_{1-x}Se$ insulating and magnetic Fe-deficient phase, exhibiting analogue Fe vacancy ordering patterns [107]. The $A_xFe_{2-y}Se_2$ compounds challenge the common picture of superconductivity related to scattering between the hole and electron pockets enhanced by Fermi surface nesting, indeed, only electron Fermi surfaces around the zone corners are observed [10]. In this respect, new scenarios of pairing mechanism and superconducting state symmetry have been proposed to reconcile the case of $A_xFe_{2-y}Se_2$ with other iron based superconductors [108].

As demonstrated in ref. [109], the electronic and magnetic phase diagram of $K_xFe_{2-y}Se_2$ system can be drawn as a function of Fe valence state, which in turns is determined by x and y. For $V_{Fe} \leq 1.935$ and $\geq 2.00$, two highly insulating phases are found, having a gap larger than 0.3 eV, antiferromagnetic order with Néel temperature as high as 559 K and ordered magnetic moment larger than $3\mu_B$. These phases also display iron vacancy order. Superconductivity occurs in a narrow region of Fe valence from ~1.94 to 2.00 between these two insulating phases. In this region, superconductivity coexists with a long range AFM order in a microscopic phase separated scenario. The value of room temperature Seebeck coefficient measured in three regions of the phase diagram plotted in the T-$V_{Fe}$ plane shows a very interesting behavior, giving important clues about the nature of charge carriers and changes in the Fermi surface across the phase diagram. In particular, as shown in the upper inset of Figure 17, the Seebeck coefficient values are large and positive or large and negative in the two insulating phases $V_{Fe}>2.00$ and $V_{Fe}<1.935$, respectively, indicating opposite types of the dominant charge carriers in these two insulating phases, while in the superconducting region, very small values of S(300K) are observed. A divergent behavior in Seebeck coefficient occurs at the boundaries between AFM insulating and superconducting phases, pointing to the existence of Lifshitz transitions. The S(T) curve of $K_{0.8}Fe_{2.2}Se_2$, whose iron valence lies in the 1.935< $V_{Fe}$ <2.00 superconducting region, exhibits a broad negative minimum around 120K, with magnitude $|S|\approx 65$ microV/K (see Figure 17), rather similar to S curves of electron doped 122 pnictides.

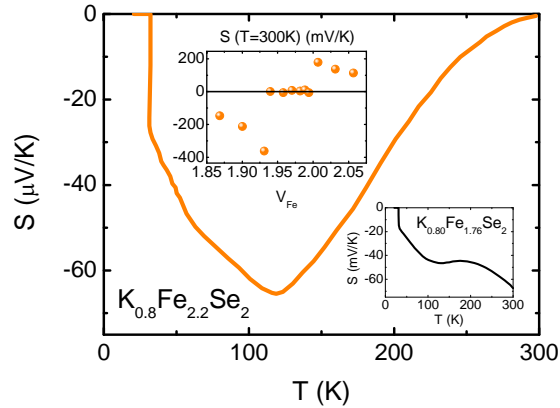

**Figure 17:** Main panel: S(T) curve of a superconducting $K_{0.8}Fe_{2.2}Se_2$ single crystal ($T_c$=31.8K). In the upper inset, the S(300K) values of $K_xFe_{2-y}Se_2$ crystals are plotted as a function of the Fe valence state $V_{Fe}$, evidencing sharply different behaviors in the three regions of the phase diagram. Data are taken from ref. [109]. In the lower right inset, the S(T) curve of a superconducting $K_{0.80}Fe_{1.76}Se_2$ single crystal ($T_c$=30K) taken from ref. [111] is also displayed, evidencing the sensitivity of S to stoichiometry, especially in the high temperature regime.

Below the S minimum, the S(T) curves of superconducting $K_xFe_{2-y}Se_2$ crystals follow a linear temperature dependence, indicating that thermopower is dominated by diffusion of n-type carriers. In ref. [27], the Seebeck curve of a single crystal with composition $K_{0.65}Fe_{1.41}Se_{2.00}$ is linearly fitted in the temperature range from $T_c$=32K to 130K. In a single picture, the linear slope yields a Fermi temperature $T_F$=880K (eq. (6)). The $T_c/T_F$ ratio, which is a measure of the correlated character of a superconductor, turns out to be ~0.04, i.e. much smaller than the value ~0.1 found in the strongly correlated $Fe_{1+y}Te_{1-x}Se_x$ [110], but smaller than the value found in low $T_c$ superconductors ~0.02. This places this compound in the weakly or intermediate correlated regime. Remarkably, the strongly insulating AFM parent compound would point to similarity with the Mott AFM insulating parent compounds of strongly correlated high-$T_c$ cuprates. However, it is suggested that the ordered Fe vacancies could induce band narrowing and consequently decrease the correlation strength needed for the Mott transition in the parent compound.

It must be underlined that 122 iron chalcogenides are intrinsically off-stoichiometric and their properties depend sensitively on the exact stoichiometry. This is clearly seen by comparing the Seebeck curve measured in a $K_{0.8}Fe_{2.2}Se_2$ single crystal with $T_c$=31.8K [109], which is displayed in the main panel of Figure 17, with the Seebeck curve measured in a $K_{0.80}Fe_{1.76}Se_{2.00}$ single crystal with $T_c$=30K [111], displayed in the lower right inset of Figure 17. The curves are pretty similar in value and shape below the S minimum around 100K, but quite different at higher temperatures. Indeed the latter S curve exhibits a local maximum around 200K with S≈-45 µV/K and a room temperature value S≈ -65 µV/K, in sharp contrast with the vanishing room temperature S of the $K_{0.8}Fe_{2.2}Se_2$ crystal in the main panel of Figure 17. This variability and non monotonic behavior is possibly related to the multiband structure of this compound and crossover between metallic and semiconducting regimes, which are sensitive to stoichiometry changes, especially Fe stoichiometry which tunes the phase diagram from the insulating Fe vacancy ordered parent compound to the metallic and superconducting phase.

## 5.4 Seebeck effect of 11 superconductors

By substituting Se on the Te site, the antiferromagnetic ordering temperature of FeTe is gradually decreased and superconductivity appears (see review [7]). In the $FeTe_{1-x}Se_x$ system, the highest $T_c$≈16K is found in $FeTe_{0.5}Se_{0.5}$, and it decreases to $T_c$≈13 K in FeSe. For x approaching the end member FeSe, apart from a limited range of phase separation 0.6<x<0.8, a tetragonal-orthorhombic structural transition appears and its temperature gradually grows with x, being observed in FeSe at

70K-100K [112,113,114]. It must be noted that (Te,Se) substitution is isoelectronic, hence in principle the charge compensated nature of the parent compound FeTe is not affected by this substitution. However, excess Fe in the +1 valence state dopes electrons into the system. As the structure is stabilized by either excess Fe or Se substitution, with increasing Se substitution, less excess Fe is required. For this reason, the effect of Se substitution results in less electron doping and is equivalent to hole doping.

A series of Seebeck curves measured in $Fe_{1+y}Te_{1-x}Se_x$ single crystals [57] is shown in Figure 18. In the high-temperature regime, the Seebeck coefficient is constant as a function of temperature and its value increases monotonically with increasing Se content x, from a negative value S(300K)≈−0.85µV/K for the y=0 sample (see Figure 8) to a positive saturation value 6-7 µV/K for the samples with x≥0.3. This constant behavior is a distinctive feature of 11 compounds and is described by the Heikes law eq. (8) as already discussed for the FeTe parent compound. The monotonic tendency to positive thermopower values with increasing Se content is consistent with the above argument that Se substitution is equivalent to hole doping. Further effects of band modifications by Se substitution are not easily evaluated, because the balance of electron and hole bands in $Fe_{1+y}Te_{1-x}Se_x$ is particularly close to compensation, as also seen from the experimental curves of Hall resistance that show multiple sign changes and do not behave monotonically with Se content [36]. The magnitude of thermopower is generally smaller than in other iron-based families, confirming the highly compensated character of electronic properties of the 11 family. As shown in Figure 18, in the $Fe_{1+y}Te_{0.9}Se_{0.1}$ sample, the feature associated to the magnetic/structural transition is shifted from 66K of $Fe_{1+y}Te$ to below 50K and is significantly broadened. In the $Fe_{1+x}Te_{0.8}Se_{0.2}$ sample the magnetic/structural transition is further suppressed. The samples with x≥0.2≤x≤0.45 exhibit similar behavior, namely from the constant high temperature value they undergo a crossover where the Seebeck changes in sign, they show a negative minimum around 18K-35K and finally, their Seebeck drops to zero at the superconducting transition temperature 11-13K.

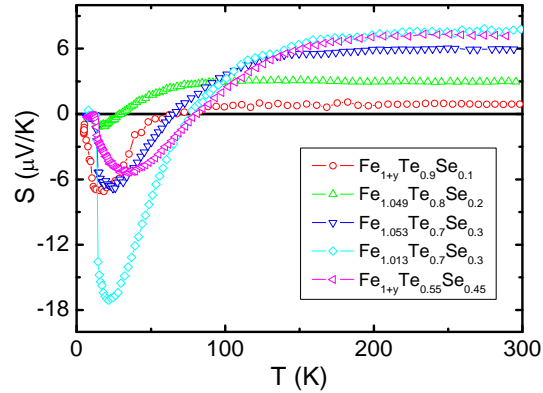

**Figure 18:** S(T) curves of $FeTe_{1-x}Se_x$ single crystals. Data are taken from ref. [57].

Negligible magnetic field dependence is detected in $Fe_{1+y}Te_{1-x}Se_x$ single crystals [57], thus ruling out any significant magnon drag contribution, similar to the case of the 11 parent compound (see section 4.3). However, the issue of coupling of charge carriers with spin fluctuations in this system must be considered and examined to understand Seebeck and transport experimental data, because spin fluctuations may scatter charge carriers even if they do not yield a magnon drag Seebeck contribution. In ref. [58], electric and thermoelectric transport properties of two superconducting single crystals of similar compositions, $Fe_{1.01}Te_{0.62}Se_{0.38}$ and $Fe_{1.01}Te_{0.62}Se_{0.40}$, and similar $T_c$ 13.4K and 13.9K respectively, are compared. The former sample, $Fe_{1.01}Te_{0.62}Se_{0.38}$, exhibits semiconducting behavior in the normal state above $T_c$, while the latter sample, $Fe_{1.01}Te_{0.62}Se_{0.40}$, exhibits metallic behavior. Although the resistivity curves of the two samples are pretty different, the corresponding S curves are very similar apart from the magnitude of S at its minimum. The authors argue that the slight increase in Se content shifts the systems from a region of coexistence of (π,π) and (π,0) fluctuations, to the region of domination of (π,π) fluctuations. Keeping in mind that (π,0) magnetic order fluctuations are antagonistic to metallicity, while (π,π) fluctuations are supposed to promote superconductivity, the authors explain the observed properties in a multiband scenario, assuming that electric and thermoelectric properties are dominated by different bands, which are differently coupled to (π,0) and (π,π) spin fluctuations, that compete with or promote superconductivity.

The generally observed proportionality between the magnitude of the negative S minimum and $T_c$ in 11 superconductors [36,57] supports a picture where both S minimum and $T_c$ are related to spin fluctuations, even if the progressive departure from the nesting compensated condition with increasing Se content could be a further reason for the proportionality between the magnitude of the negative S minimum and $T_c$.

Despite the magnitude of S in superconducting 11 compounds in generally smaller than in other iron-based superconducting families, in ref. [110] it is noted that $Fe_{1+y}Te_{0.6}Se_{0.4}$ has the largest |S| among Fe(Te,Se), up to 38 microV/K, which could be a direct consequence of strong electronic correlations leading to a reduced Fermi energy. The Fermi temperature, $T_F$, deduced from low temperature S data in a single band approximation, yields a large $T_c/T_F$ ratio comparable to other correlated superconductors whose pairing is mediated by spin fluctuations and places Fe(Te,Se) among strongly correlated superconductors.

We now focus on the end member of the series $FeSe_{1-y}$, where superconductivity appears upon suppression of the Fe vacancy ordered phase of the Fe deficient insulating and magnetic parent compound [107], rather than from the FeTe parent compound. As for Fe(Te,Se), also $FeSe_{1-y}$ shows extreme sensitivity of superconducting and electric transport properties to tiny changes in stoichiometry, associated to Se vacancies [114]. In ref. [67], it is shown that when the compound is closest to exact stoichiometry $T_c$ is 8.5K, but it drops to 5K in $FeSe_{0.98}$ and is non superconducting down to 0.6 K in $FeSe_{0.97}$. Despite the remarkable effect of stoichiometry on $T_c$ and on the shape of the resistivity curve, the Seebeck coefficients are qualitatively similar for superconducting and non superconducting $FeSe_{1-y}$ (see Figure 19), namely S is small and positive at room temperature, changes sign around 200-230 K, goes through a broad negative minimum near the structural phase transition around 90-110K and changes sign again around 20-30K, in correspondence of the change of curvature of the S-shaped resistivity curve [67,69,115]. Clearly a competition of almost compensated hole and electron bands comes into play. Regarding the common behavior of Seebeck curves of different $FeSe_{1-y}$ samples, a multiband picture as the one proposed in ref. [58] can be hypothesized.

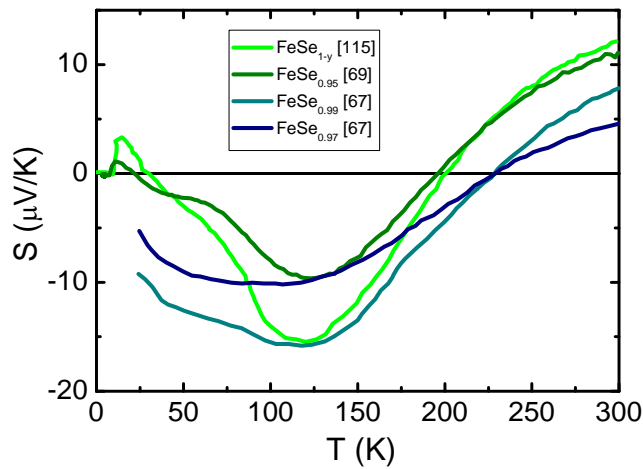

**Figure 19:** S(T) curves of $FeSe_{1-y}$ polycrystals with slightly different stoichiometry taken from different works, namely almost stoichiometric $FeSe_{1-y}$ with $T_c$=9.4K [115], $FeSe_{0.95}$ with $T_c$=8K [69], $FeSe_{0.99}$ with $T_c$=9K [67], $FeSe_{0.97}$ with no $T_c$ down to 0.6K [67].

The effect of doping on superconductivity and thermoelectric properties may add further information on the electronic structure and the balance between hole and electron bands in FeSe. In ref. [116] the effect of Cobalt doping in FeSe is explored. Superconductivity appears to be destroyed by as little as 5% Co substitution and typical metallic behavior is observed for all doping levels up to 50%. S is negative in the whole temperature range below room temperature for x≥5%, indicating electron doping by Co. The room temperature magnitude of the Seebeck coefficient |S(300K)| increases with increasing Co content x up to 25%, likely due to progressive departure from a compensated character. All the curves for 0≤x≤0.25 exhibit a broad minimum, whose largest absolute value is ~80 µV/K at around 100K for x=5%. Hence Co enhances S, which, combined with rather low resistivity $\rho$, yields considerable power factors $S^2/\rho$ larger than 15 and figures of merit $ZT=TS^2/(\rho\kappa)$ in the range $10^{-2}$. In ref. [117] electron doping in FeSe is obtained by to substitution of trivalent Bi for divalent Fe. $T_c$ decreases slightly with increasing Bi content, from 10.02K to 8.26K for Bi content from 0 to 8%. Thermopower measured up to 700K shows a double sign crossing above and below room temperature. This fact, joined with the low magnitude of S in the temperature range 200K-700K, indicates that both holes and electrons contribute significantly to transport in the high temperature regime.

**5.5 Seebeck effect of 111 superconductors**
Very few Seebeck curves can be found in literature for superconducting compounds of the 111 family. We mention the data measured on LiFeAs [118], where negative S values below 100K are shown down to the superconducting transition at $T_c$=18K, and data measured on $Na_xFeAs$ ($T_c$=12-25K) [119], which closely resemble the Seebeck curve of almost stoichiometric $FeSe_{1-y}$ [115] displayed in Figure 19.

**6. Comparison of between Seebeck behaviors in parent and superconducting compounds of the different families**
In this section, the different phenomenologies observed in various iron-based compounds are summarized, evidencing peculiar and common behaviors.
Starting from the parent compounds, in all the families a sharp jump of S(T) at the structural/magnetic transition is observed, likely related to a reconstruction of the Fermi surface. In 1111 and 122 parent compounds, S jumps to more positive (less negative) values, while in the 11 parent compound it jumps to more negative values, indicating a different rearrangement of the balance between hole and electron sheets from the paramagnetic to the antiferromagnetic state. In all the families, the room temperature value of S of the parent compounds is very small, typically |S|~10 µV/K or smaller, indicating strong electron-hole compensation. The sign of S is negative for 1111 and 11 parent compounds, while for some of the 122 parent compounds it is positive, as in the cases of $CaFe_2As_2$ and $EuFe_2As_2$. Both 1111 and 122 parent compounds are characterized by strong variability of Seebeck behavior at temperatures below the structural/magnetic transition from sample to sample, even with identical composition. This lack of reproducibility can be ascribed to the disorder easily incorporated in the complex crystal structure, which is highly effective in suppressing the magnon drag contribution to S, represented by a broad negative peak around 50K. The Seebeck behaviors of the 122 parent compounds present also systematic changes in dependence of the alkaline earth metal, either Ba, Ca or Eu, which can be described as almost rigid shifts of the S curves and can be attributed to a different hole-electron balancing in the transport contributions. Specifically, smaller ionic radius of the alkaline earth metal seems to be correlated with larger hole contribution to transport. In general, the non monotonic S(T) behavior below the structural/magnetic transition, characterized by broad peaks and changes of sign, is common to 1111 and 122 parent compounds, and it is determined by the presence of multiband diffusive and drag contributions to S. In this respect the 11 parent compound is a bit different, in that it exhibits no evidence of magnon drag Seebeck, being S insensitive to application of magnetic field. Moreover, the Seebeck of the 11 parent compound is characterized by a strikingly flat temperature dependence above the structural/magnetic transition. This Heikes regime achieved at fairly low temperature is related to strong electronic correlations.

Regarding the superconducting compounds, a common phenomenology can be identified for all the families. As doping gradually decreases the structural/magnetic transition temperature and eventually superconductivity appears, the shape of the Seebeck curves changes accordingly, with the feature associated to the structural/magnetic transition disappearing and the superconducting transition appearing. The Seebeck curves present either a broad negative minimum (for electron doping) or positive maximum (for holes doping) around 100-150K for 1111 and 122 superconductors and around 18K-35K for 11 superconductors. Superconductivity with hole doping occurs only in the 122 family, where positive Seebeck curves are observed. Both in 1111 and 122 families, pretty large |S| in the superconducting compounds are obtained, even exceeding 100µV/K, and it is reasonable to assume that common reasons lie at the origin of such enhancement of S magnitude, possibly spin fluctuations, strong electron correlation, peculiar electronic structure. However taking some key features into account such as the low carrier concentration semimetal character, the moderately high density of states originated primarily from Fe $d$ bands, the compensating electron and hole Fermi surfaces and the very small Fermi energy of the undoped parent compounds, diffusive |S| around 50-90 µV/K can be predicted [94]. On the other hand, 11 superconductors have smaller |S|, few tens µV/K at most, possibly due to larger compensation of electron and hole bands. Both in 1111 and 122 families, a non monotonic behavior of S as a function of doping is observed, indicating departure from electron-hole compensation at low doping and increasing carrier concentration of one type at high doping. Remarkably, in 122 superconductors close to optimal doping, evidence of critical fluctuations is observed from the logarithmic temperature dependence of S/T in proximity of the QCP, $S/T \propto \ln(1/T)$. Finally, as in the case of the 11 parent compound, also 11 superconductors reach the flat Heikes regime at temperatures as low as 100-200K, indicating a significant role of electron correlations.

## 7. Nernst effect of parent and superconducting compounds of iron pnictides and chalcogenides

As compared to the Seebeck effect, the Nernst effect is more difficult to measure, due to its small magnitude, necessity of high fields and necessity of subtraction of spurious contributions from the signal. Also the data interpretation is more complex, due to the presence of multiple alternative or coexisting mechanisms potentially into play, related to band structure, superconducting mechanisms and fluctuations. These reasons explain why a limited amount of Nernst data is available in literature. Moreover a different sign definition of the Nernst signal among literature data must be taken into account.

### 7.1 Nernst effect of compounds of the 1111 family

In the top panel of Figure 20, the Nernst coefficients ν measured on LaFeAsO polycrystals, taken from ref. [44] and measured on the same sample whose Seebeck is reported in ref. [18], are plotted as a function of temperature. For both samples, the Nernst coefficient is small above $T_{SDW}$, undergoes an abrupt increase in magnitude at $T_{SDW}$ toward positive values, exhibits a maximum around 100K, a change in sign below 50K and a negative minimum around 10-20K. The magnitude of ν reaches values around ν≈0.5 µVK$^{-1}$T$^{-1}$, remarkably larger than typical values of ordinary metals, but not extraordinarily enhanced as in bismuth and heavy-fermion metals [21]. A severe violation of the Sondheimer cancellation in the SDW phase is detected for both samples. More interestingly, the Nernst signal $e_y$ measured up to 8-9T for the sample of ref [44] and up to 30T the sample of ref. [18] is linear in the temperature range where the Nernst signal is positive and departs visibly from linearity in the negative low temperature range. This is shown in the inset of Figure 20 for the sample of ref. [18]. The same temperature trend is confirmed also by measurements carried out on the parent compound $RE$FeAsO ($RE$=La, Sm, Pr, Ce) polycrystals of ref. [18], all showing anomalous large positive Nernst coefficients below $T_{SDW}$, reaching values up to $\nu \approx 0.5$ µVK$^{-1}$T$^{-1}$, as displayed in the bottom panel of Figure 20. These data also indicate that at low temperature T<50K, the thermoelectric behavior changes from sample to sample, with a negative Nernst contribution that is either well evident or absent. A similar erratic behavior at low temperature is also observed in the Seebeck curves (see section 4.1), as a consequence of disorder and its effect on the magnon drag. Apparently, Seebeck and Nernst curves are determined by the same mechanisms and a correspondence exists between them. The observation of the non linear field dependence of the negative Nernst term (inset of Figure 20) is a further clue pointing to the coupling with magnon excitations as the mechanism responsible for the negative Nernst signal at low temperature in these compounds. In the high temperature regime, qualitatively, the sharp change of behavior of the Nernst coefficient at $T_{SDW}$ could be related to the Fermi surface reconstruction. Concerning the mechanisms for the enhancement of ν below $T_{SDW}$, authors of ref. [44] suggest that spin-dependent scattering processes related to SDW order or SDW fluctuations, which could be band-dependent, could play a role, in analogy with the Nernst effect of the $p$-wave superconductor Sr$_2$RuO$_4$, explained in terms of changes in scattering mechanisms [120]. Another possible alternative or additional cause for the enhancement of the Nernst coefficient is the presence of a band with Dirac dispersion crossing the Fermi level, which applies to 122 [24] and 1111 [121] iron pnictide parent compounds alike. It was pointed [23] out that band degeneracy in the SDW state originates gapless nodal points along the Fermi surface. In particular, the Fermi surfaces connected by the SDW wave vector have a vorticity mismatch that leads to a nodal SDW and creates Dirac cones near the Fermi energy. A confirmation of the presence of bands with linear dispersion at the Fermi level is given by the observation of linear positive magnetoresistance in 1111 [122] parent compounds and linear positive magnetoresistance [123] and ARPES [124] measurements in 122 ones. This peculiar situation of Dirac bands is addressed theoretically in ref. [125]. The authors derive an effective Hamiltonian for Dirac fermions and build a phenomenological two-band model consisting of a hole band with a conventional energy spectrum and an electron band with Dirac energy spectrum. As a consequence of the high mobility of Dirac electrons, some transport properties are dominated by the contribution of Dirac fermions, even if the carrier concentration of Dirac fermions is much smaller than that of conventional carriers [122]. Remarkably, the model predicts a large and positive contribution from the Dirac fermions for the Nernst coefficient, because the quantity is strongly dependent on the mobility (see section 3.3). The observed linear field dependence of the Nernst signal in the temperature range of several tens K below $T_{SDW}$ is consistent with the Dirac fermions scenario, as long as $\omega_c \cdot \tau < 1$.

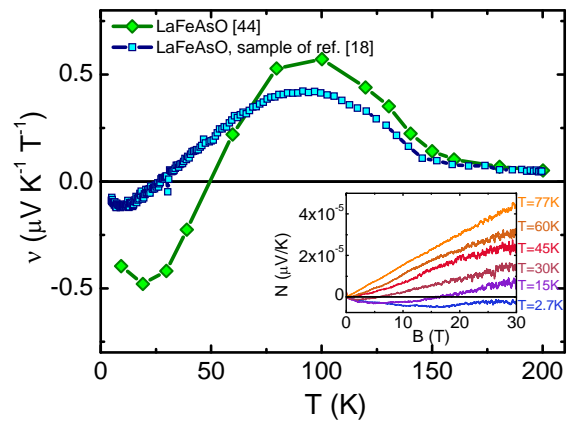

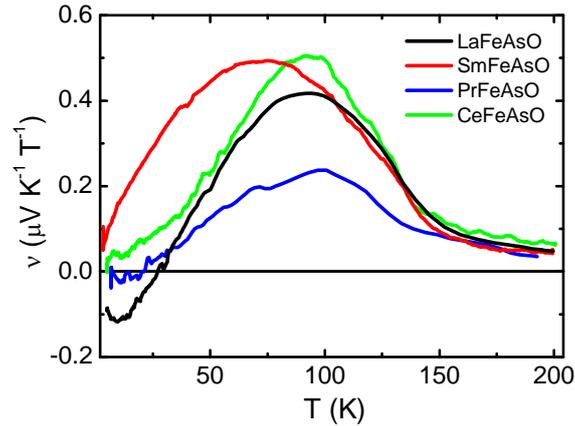

**Figure 20:** Top panel: Nernst coefficient ν(T) curves measured in parent compound LaFeAsO polycrystals. Data are taken from ref. [44] and from Nernst measurements carried out on the sample of ref. [18]. In the inset, the Nernst signal N of the sample of ref. [18] is plotted as a function of field up to 30T. Bottom panel: ν(T) curves measured in the parent compound REFeAsO (RE=La, Sm, Pr, Ce) polycrystals of ref. [18].

In ref. [44], an interesting comparison is carried out between Nernst responses of LaFeAsO and LaNiAsO. Low-$T_c$ superconductivity, likely of conventional type, is observed in NiAs(P)-based compounds with similar layered structure as FeAs-based compounds, but with neither structural transition nor AFM ordering. At odds with the enhanced Nernst effect in LaFeAsO below $T_{SDW}$, the Nernst effect in LaNiAsO is very small ($e_y$ at B=6 T is in the range of ±30 nV/K, almost two order of magnitude smaller than that of LaFeAsO) and weakly temperature dependent, and the Sondheimer cancellation is partially held, as in normal metals. The authors suggest that high-$T_c$ unconventional superconductivity and anomalous thermoelectric properties are related in FeAs-based systems.

Nernst coefficients ν measured on several superconducting LaFeAsO$_{1-x}$F$_x$ polycrystals are shown in Figure 21, namely an underdoped sample, LaFeAsO$_{0.95}$F$_{0.05}$ ($T_c$=20.6K) taken from ref. [32] and two optimally doped LaO$_{0.9}$F$_{0.1}$FeAs samples ($T_c$≈26 K) taken from ref. [32,126]. A strong positive contribution arising from the vortex motion in the superconducting state is present in all samples, as emphasized in the inset of Figure 21. As expected, vortex Nernst signal is nonlinear with magnetic field in all the samples. Below ~10 K, vortices are pinned so that the vortex Nernst signal vanishes, while above ~25K the vortex Nernst signal is no longer dominating, as the vortex lattice is progressively destroyed by thermal fluctuations. This vortex liquid state regime occurs in a quite large temperature window, similar to the case of high $T_c$ cuprates and much larger than the case of conventional type-II superconductors. Despite very similar behaviors of optimally doped and underdoped samples in the vicinity of $T_c$, the Nernst signals are different in the normal state. For the underdoped sample LaFeAsO$_{0.95}$F$_{0.05}$, in the normal state, ν(T) is positive and rather flat between 300 K and 150 K, it changes in sign at ~100 K and presents a minimum at ~40 K, below which the above mentioned vortex Nernst contribution onsets. A significant violation of the Sondheimer cancellation is seen as in the parent compound LaFeAsO. However, the magnitude of ν in LaFeAsO$_{0.95}$F$_{0.05}$ is more than one order of magnitude lower than in LaFeAsO. In the optimally doped samples LaO$_{0.9}$F$_{0.1}$FeAs just above $T_c$≈26 K the Nernst signal is negative, small, weakly temperature dependent and linear as a function of magnetic field. Despite the temperature dependence is weak in the normal state, it can be noted that around 50 K, ν(T) starts to deviate from the negative background and increases gently. Such behavior occurs in cuprates as well, where it was interpreted in terms of vortex excitations above $T_c$. However, the authors of ref. [126] provide an alternative picture. They analyze separately the two terms of the Sondheimer cancellation by measuring the Hall effect and Seebeck effect and find out that cancellation is not complete (in any case the violation of the Sondheimer cancellation is milder than in the underdoped sample, as expected from the more metallic properties at optimal doping). This is a consequence not so much of the multiband character, but rather of the anomalous temperature behavior of the normal state off-diagonal Peltier coefficient term $\rho \cdot \alpha_{nxy}$. This term should start decreasing at $T_c$ and vanish below $T_c$, however in facts it starts decreasing at 50 K, that is far above $T_c$, which means a change in the electron state at 50 K. The authors suggest that the slow increase in $\nu(T)$ above 50 K might be caused by the residual magnetic fluctuations from the SDW state of the parent compound suppressed by F doping, and accordingly the sharp decrease in $\rho \cdot \alpha_{nxy}$ below 50 K results from the suppression of these magnetic fluctuations, which preludes the occurrence of superconductivity at $T_c$. The authors suggest that in LaFeAsO$_{1-x}$F$_x$ SDW and superconducting order parameters do not coexist but truly compete with each other, at odds with 122 systems such as Ba(Fe$_{1-x}$Co$_x$)$_2$As$_2$, where coexistence of static magnetism and superconductivity is evidenced in certain regions of the phase diagram.

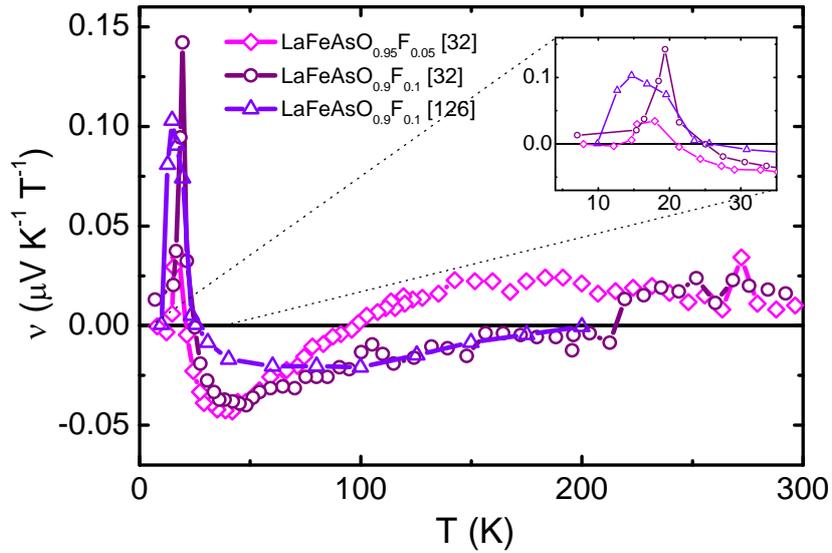

**Figure 21:** $\nu(T)$ curves measured in superconducting LaFeAsO$_{1-x}$F$_x$ polycrystals with x=0.05 [32] and x=0.1 [32,126]. In the inset, a magnification of the low temperature vortex regime is shown.

Given that in conventional metals the magnitude of the Nernst effect is usually in the nVK$^{-1}$T$^{-1}$ range [21], the enhanced Nernst effect in parent compounds and underdoped high-$T_c$ cuprates and iron pnictides in the hundreds nVK$^{-1}$T$^{-1}$ range is surprising and could suggest a common relationship between high Nernst effect and unconventional superconductivity. A comparison between the Nernst effect in LaFeAsO$_{1-x}$F$_x$ and in high-$T_c$ cuprates La$_{1.8-x}$Eu$_{0.2}$Sr$_x$CuO$_4$ and La$_{2-x}$Sr$_x$CuO$_4$ is carried out in ref. [127]. In cuprates static stripe order and fluctuating stripes may be responsible for the enhanced Nernst response [128]. Unusual enhancement of the Nernst coefficient in the normal state of cuprate high $T_c$ superconductors at temperatures much higher than the critical temperature $T_c$ is also interpreted as the signature of vortex fluctuations [28]. On the other hand, as discussed above in this section, in the iron pnictides Fermi surface reconstructions related to formation ordered phases, specifically formation of Fermi surface pockets, possibly with Dirac dispersion, could yield enhancement of the Nernst coefficient in the parent compounds. In the normal state of the iron pnictide superconducting compounds, the Nernst signal may be enhanced by fluctuating forms of the SDW ordered state, as the Nernst signal may sensitively detect the entropy associated to spin fluctuations.

### 7.2 Nernst effect of compounds of the 122 family
In Figure 22, the Nernst coefficients $\nu$ measured on single crystals of 122 parent compounds, namely EuFe$_2$As$_2$ and CaFe$_2$As$_2$, taken from ref. [50] and [48] respectively, are shown.

In EuFe$_2$As$_2$ [50], ν is positive in the whole temperature range, undergoes a jump at T$_{SDW}$=191K, which correlates with the sharp increase in absolute value of S, and becomes anomalously large below T$_{SDW}$, reaching around 600 nVK$^{-1}$T$^{-1}$. A strong departure from the Sondheimer cancellation is observed. The properties of EuFe$_2$As$_2$ exhibit signatures of Dirac fermions at the Fermi surface. Indeed, experimental curves of Hall resistance R$_H$, S and ν are well reproduced by the phenomenological two-band model developed in ref. [125], where a hole band with conventional energy spectrum, and an electron band with linear energy spectrum are assumed and R$_H$, S and ν are calculated from the elements of the Peltier and electrical conductivity tensors. A pretty large low-temperature mobility ~98 cm$^2$V$^{-1}$s$^{-1}$ is extracted, which can be a manifestation of the small effective mass of the Dirac fermions. Moreover, at low temperature (T<10K), the ν data depend linearly on the temperature, so that if eq. (13) is used to extract the Fermi energy, a value E$_F$≈7 meV is obtained. This low Fermi energy value suggests the proximity of the Fermi level to the Dirac cone vertex in this sample, which could account for the anomalous enhancement of ν. The Nernst coefficient in CaFe$_2$As$_2$ also shows a sharp increase below T$_{SDW}$, and an overall behavior qualitatively very similar to EuFe$_2$As$_2$, so that all the above considerations likely apply to the case of CaFe$_2$As$_2$, where E$_F$≈20 meV is obtained from eq. (13). We remark that these EuFe$_2$As$_2$ and CaFe$_2$As$_2$ samples do not exhibit the low temperature Nernst contribution observed in 1111 parent compounds (see of Figure 20), tentatively associated to the coupling with magnon excitations. Consistently, in the Seebeck curves measured in the same samples (see Figure 5), the negative minimum around 20K attributed to magnon drag is significantly suppressed as compared to the case of some 1111 parent compounds, exhibiting a large magnon drag minimum around 50K (see Figure 2).

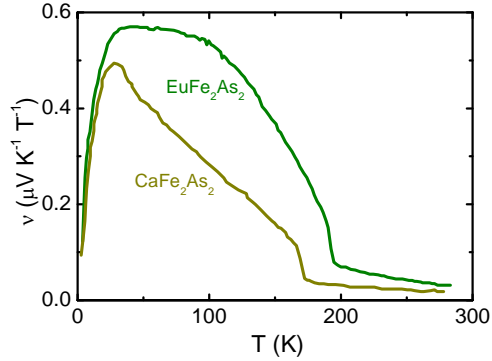

**Figure 22:** ν(T) curves measured in 122 parent compounds EuFe$_2$As$_2$ [50] and CaFe$_2$As$_2$ [48] single crystals.

In the same works, ref. [50] and [48], the Nernst coefficients of the corresponding doped superconducting compounds are measured and the curves are shown in Figure 23. In ref. [50], the Eu(Fe$_{1-x}$Co$_x$)$_2$As$_2$ series of single crystals with x=0, 0.15, 0.20 and 0.30 is studied. In the x=0.15 sample where SDW is still present below 131K, there is still a certain enhancement of ν. With increasing cobalt doping the SDW transition disappears, the system shifts towards the characteristics of a regular metal, represented by the x=0.30 sample, namely R$_H$ is small and weakly temperature dependent, S is nearly linear with the temperature and ν becomes very small (|ν|<5 nV K$^{-1}$ T$^{-1}$) as expected in the case of the satisfied Sondheimer cancellation. The authors attempt to separate the anomalous (related to SDW) and normal (expected from a regular metal) contributions to the Nernst signal. The dramatic effect of Co doping on the Nernst signal must be related to the disappearance of the SDW state and of the influence of the Dirac fermions. Indeed, the Fermi energy extracted from the low-temperature Nernst data systematically increase with increasing Co doping, suggesting that the Fermi level steadily departs from the Dirac cone vertex. Again, similar considerations applies to the case of CaFe$_{1.92}$Co$_{0.08}$As$_2$ [48], also shown in Figure 23.

Remarkably, no detectable vortices contribution to the Nernst signal in the superconducting state is observed, either in Eu(Fe$_{1-x}$Co$_x$)$_2$As$_2$ or in CaFe$_{1.92}$Co$_{0.08}$As$_2$. The reason is due to the narrow temperature range where it appears, explained by strong vortex pinning and reduced thermal fluctuations in these materials [129].

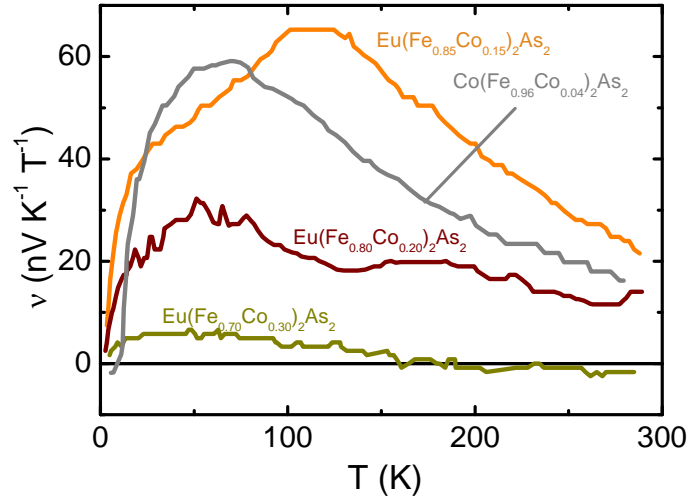

**Figure 23:** $\nu(T)$ curves measured in 122 superconducting compounds $Eu(Fe_{1-x}Co_x)_2As_2$ [50] and $CaFe_{1.92}Co_{0.08}As_2$ [48] single crystals.

### 7.3 Nernst effect of compounds of the 11 family

The Nernst response of $Fe_{1+y}Te$ presents sharp differences as compared to those of iron-based pnictide parent compounds of 1111 and 122 families. The Nernst coefficient $\nu$, measured in ref. [58] and plotted in Figure 24, is positive, weakly temperature dependent and small in magnitude (tens nm $K^{-1}$ $T^{-1}$) above $T_N$=60K, it undergoes an abrupt jump at $T_N$, and it becomes negative below $T_N$, exhibiting a minimum around 25K. Its magnitude is smaller than those of 1111 and 122 parent compounds by more than one order of magnitude. The smaller magnitude of $\nu$ in $Fe_{1+y}Te$ can be explained by the absence of Dirac cones in $Fe_{1+y}Te$, which are likely responsible for the Nernst enhancement in 1111 and 122 parent compounds. Indeed, no evidence of bands with Dirac dispersion crossing the Fermi level has ever been observed experimentally, nor predicted by ab initio band calculations [69]. Authors of ref. [58] point out that the reconstruction of the Fermi surface due to formation of the antiferromagnetic order at $T_N$ yields the formation of spin stripes, which are indeed expected to produce a maximum (or minimum) in the Nernst signal at $T \approx 1/3 \cdot T_N \approx 25K$ [128], as observed. The Nernst signal $e_y$ shows no deviation from linearity with magnetic field.

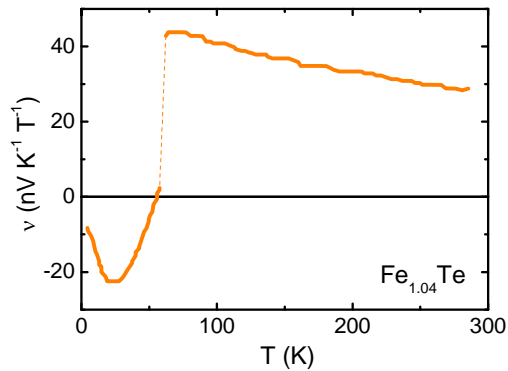

**Figure 24:** $\nu(T)$ curve measured in 11 parent compound $Fe_{1.04}Te$ single crystal. Data are taken from ref. [58].

The evolution of the Nernst coefficient with increasing Se substitution in a series of polycrystalline $Fe_{1+y}Te_{1-x}Se_x$ samples is reported in the top panel of Figure 25 (the corresponding Seebeck curves measured in the same samples are shown in Figure 7 of ref. [36]). Oppositely to the case of 1111 and 122 families, with increasing departure from the antiferromagnetically ordered sample $Fe_{1+y}Te$, the ν turns to positive values and progressively rises in magnitude, reaching values close to 200 nV/KT in the x=1 FeSe sample, not too far from the values reported for the 1111 and 122 parent compounds in the AFM state. It must be noted that the presence of Dirac cones in the band structure of FeSe was suggested on the basis of magnetotransport measurements [130] and *ab initio* calculations [131], hence the observed sizeable Nernst effect may be explained in this scenario.

The Nernst coefficients of superconducting $Fe_{1+y}Te_{1-x}Se_x$ x≈0.4 single crystals, taken from ref. [58], are displayed in the bottom panel of Figure 25. These two samples of composition $Fe_{1.01}Te_{0.62}Se_{0.38}$ and $Fe_{1.01}Te_{0.62}Se_{0.40}$, already described in section 5.4, exhibit similar $T_c$'s 13.4K and 13.9K, respectively, different transport behaviors in the normal state, namely semiconducting and metallic, respectively, and similar Seebeck behaviors. As for the Seebeck curves, also the Nernst curves of these samples in the normal state look similar in the entire temperature range. Values of ν are positive at the room temperature and exhibit a sudden drop below T ≈ 60-70 K and a change to negative sign at 20-25K. This change of sign above $T_c$ is not an intrinsic feature, being observed neither in the samples shown in the top panel of Figure 25, nor in the crystal shown in Figure 26, indicating that it is not an ever present feature. In the superconducting state, the Nernst responses of the two samples depart, indeed no contribution from the vortex motion to the Nernst effect is detected in the $Fe_{1.01}Te_{0.62}Se_{0.38}$ sample, while in the $Fe_{1.01}Te_{0.62}Se_{0.40}$ sample there is a positive peak at $T_c$, clearly identified as a vortex Nernst effect. This suggests that only the latter sample with dominant (π,π) spin fluctuations (see section 5.4) has a well-established non filamentary superconductivity. The vortex Nernst effect is more deeply investigated in ref. [110], where it is measured up to 28T in a $Fe_{1+y}Te_{0.6}Se_{0.4}$ single crystal ($T_c$≈14K). As shown in Figure 26, a significant positive vortex liquid Nernst develops around $T_c$ in a wide temperature range and vanishes at low temperature with the solidification of the vortex lattice. With increasing magnetic field, the maximum vortex liquid Nernst steadily increases up to 24 T and begins to decrease at higher fields, when the overlap between the vortex cores overwhelms the increase in the number of vortices. In addition, with increasing magnetic field, the maximum vortex liquid Nernst shifts to lower temperature. In a wide window of magnetic field and temperature, well above $T_c$, the Nernst response is enhanced above its normal-state value, with no sharp transition separating the vortex liquid and the normal state. The authors [110] attribute this normal state enhanced Nernst response to thermally-induced vortex fluctuations, as in high-$T_c$ cuprates. Indeed, from the analysis of upper critical filed and resistivity data of the $Fe_{1+y}Te_{0.6}Se_{0.4}$ crystal in a single band approximation, the authors identify this compound in its normal state as a correlated metal with a low density of heavy quasi-particles and in its superconducting state as a barely clean compound, that is with a mean-free-path slightly exceeding the superconducting coherence length.

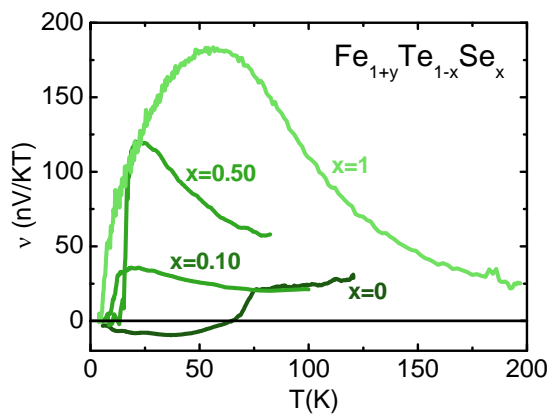

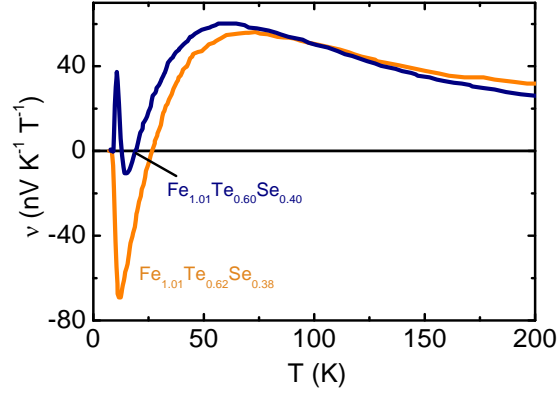

**Figure 25:** Top panel: ν(T) curves measured at 9T in the $Fe_{1+y}Te_{1-x}Se_x$ polycrystals of ref. [36]. Bottom panel: ν(T) curves measured at 12.5T in 11 superconducting $Fe_{1.01}Te_{0.62}Se_{0.38}$ and $Fe_{1.01}Te_{0.62}Se_{0.40}$ single crystals, exhibiting similar $T_c$'s 13.4K and 13.9K, respectively, and different transport behaviors in the normal state, namely semiconducting and metallic, respectively. Data are taken from ref. [58].

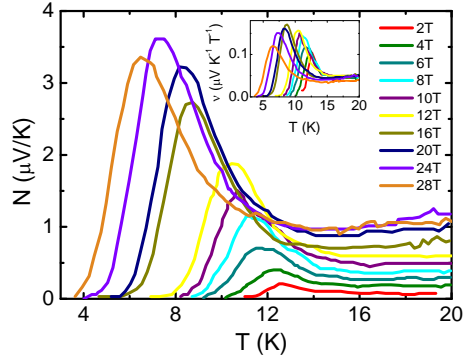

**Figure 26:** Vortex Nernst signal measured in a $Fe_{1+y}Te_{0.6}Se_{0.4}$ single crystal at different fields. In the inset, the corresponding Nernst coefficient ν=N/B is plotted. Data are taken from ref. [110].

## 8. Comparison of between Nernst behaviors in parent and superconducting compounds of the different families

Iron based compounds of the 1111, 122 and 11 families are all characterized by sizeable Nernst effect, especially the 1111 and 122 parents compounds which exhibit ν values in the in the hundreds $nVK^{-1}T^{-1}$ range, as compared to the $nVK^{-1}T^{-1}$ range of conventional metals. In general, The Nernst effect is enhanced when the Sondheimer cancellation is violated, which occurs (i) when multiple bands contribute to the transport, (ii) in presence of energy dependent Hall angle, combined with high mobility and/or low Fermi level, (iii) in case of linear dispersion of one band [22]. The latter mechanism yields a much larger enhancement of ν than the other mechanisms. It is likely that the Nernst effect is enhanced in parent compounds of the 1111 and 122 families below $T_{SDW}$ due to Fermi surface reconstructions and formation of Fermi surface pockets (low Fermi level) as well as multiband transport, however a further enhancement must be related to the formation of Fermi surface sheets with Dirac dispersion below $T_{SDW}$. By converse, in the 11 parent compound small negative ν values are observed below $T_N$, in the tens $nVK^{-1}T^{-1}$ range, consistent with the absence of Dirac cones at the Fermi surface. Also the magnitudes of Nernst effect in superconducting compounds of the 1111 and 122 families are remarkable with respect to conventional metals, especially in the underdoped regime, being typically in the tens $nVK^{-1}T^{-1}$ range. These values decrease with increasing doping. Fluctuating precursors of the SDW order and increasing departure of the Fermi level from the Dirac cone vertex are thought to play a role in determining the Nernst effect in the normal state of these superconducting compounds. On the contrary, in the 11 compounds a trend of increasing ν magnitude with increasing departure from the AFM ordered compound is observed.

A sizeable vortex signal is observed in the superconducting state of 1111 and 11 compounds whereas no detectable vortex contribution is observed in the 122 compounds. Among 1111 and 11 families, it is interesting to compare the width of the temperature windows where the vortex signal is not negligible. To this aim, in Figure 27, the resistivity transitions in Nernst curves in different magnetic fields are plotted as a function of the reduced temperature $T/T_c$, for a $SmFeAsO_{0.85}F_{0.15}$ polycrystal ($T_c \approx 52$ K) and for the Fe(Te,Se) thin film of ref. [132] ($T_c \approx 20.5$ K). The temperature range where resistivity and Nernst signal are finite, delimited by the shaded area, corresponds to the vortex liquid regime, where dissipation due to unpinned vortex motion occurs. It is clear that in both cases this temperature windows is pretty large, but in the 11 sample it is even larger than in the 1111 sample, ranging from below $T/T_c \approx 0.4$ to 1 in the former and from $T/T_c \approx 0.6$ to 1 in the latter. Extended vortex liquid regime indicates strong thermal fluctuations, which can be parameterized by the Ginzburg number Gi, defined as $Gi=(\pi\lambda_{ab}^2 k_B T_c \mu_0/2\xi_c \Phi_0^2)^2$, with $\Phi_0$ magnetic flux quantum, $\lambda_{ab}$ London penetration depth along ab planes, $\xi_c$ coherence length along the c axis, $\mu_0$ vacuum magnetic permeability. Values $Gi \approx 10^{-3}$ and $4\cdot10^{-4}$, similar as in cuprates, can be estimated for the 11 and 1111 families, respectively [129], confirming stronger thermal fluctuations in the 11 family, whose larger Gi is mainly determined by the larger $\lambda_{ab}$. For the 122 family, a smaller $Gi \approx 1.5\cdot10^{-5}$ [129] indicates a minor effect of thermal fluctuations, consistent with the absence of any detectable vortex Nernst signal.

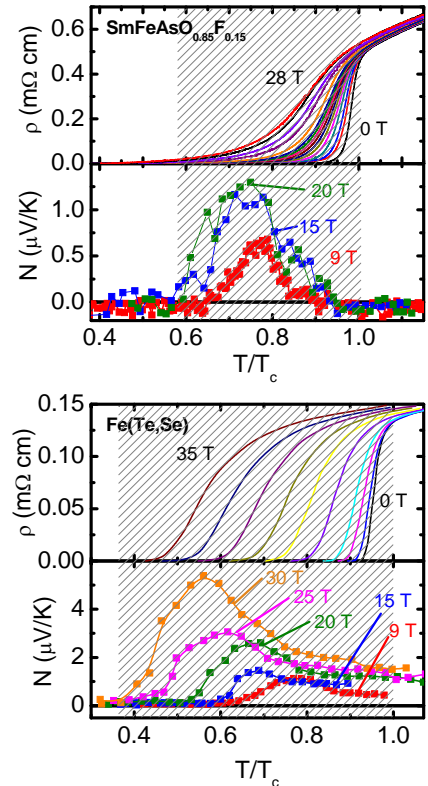

**Figure 27**: Resistive transitions and Nernst curves in different magnetic fields are plotted as a function of the reduced temperature $T/T_c$, for a $SmFeAsO_{0.85}F_{0.15}$ polycrystal (sample from ref. [133]) and a Fe(Te,Se) thin film (data from ref. [132]). The shaded area indicates the vortex liquid regime.

## 9. Conclusions

In this review, the phenomenological behaviors of Seebeck and Nernst effects of iron-based superconductors and parent compounds, extracted from literature data and from original experimental data, are collected and analyzed, in the effort of highlighting the differences and analogies between the three main families, namely 1111, 122 and 11. The main results and the comparison between families are summarized in section 6 for Seebeck data and in section 8 for Nernst data, respectively. From the Seebeck effect data on the parent compounds, information is extracted about Fermi surface reconstruction and Lifshitz transitions (1111, 122 and 11), multiband character (1111, 122 and 11), coupling of charge carriers with spin excitations and its relevance in the unconventional superconducting pairing mechanism (1111 and 122), nematicity (122), correlation (11). The Seebeck effect of superconducting compounds gives evidence of quantum critical fluctuations at optimal doping (122), correlation (11), multiband character (1111, 122 and 11). The Nernst effect has been less investigated, however its large magnitude in the parent compounds indicates the presence of Dirac dispersion bands at the Fermi level (1111 and 122), multiband transport and low Fermi level (1111, 122 and 11), while in the superconducting compounds fluctuating precursors of the spin density wave state are thought to play the major role (1111, 122 and 11).

Despite the complexity of these mysterious and promising materials, it turns out that the exploration of thermoelectric properties provides precious clues toward complete understanding of the physical mechanisms into play, eventually responsible or closely related to unconventional high $T_c$ superconductivity.

**Acknowledgements**
The authors acknowledge G. Lamura, M. Meinero and M. Tropeano for their contribution in the measurements, as well as A. Jost and U. Zeitler for high magnetic field measurements at HFMLRU/FOM. Financial support from the PRIN project No. 2012X3YFZ2 is also acknowledged.

**References**

[1] J.G. Bednorz, K. A. Muller, Z Phys B 64, 189 (1986)
[2] K. Kamihara, Y. Kamihara, T. Watanabe, M. Hirano, H. Hosono, J. Am. Chem. Soc. 130, 3296 (2008)
[3] Z.-A. Ren, W. Lu, J. Yang, W. Yi, X.-L. Shen, Z.-C., G.-C. Che, X.-L. Dong, L.-L. Sun, F. Zhou, Z.-X. Zhao, Chin. Phys. Lett. 25, 2215 (2008)
[4] M. Fujioka, S. J. Denholme, M. Tanaka, H. Takeya, T. Yamaguchi, Y. Takano, Appl. Phys. Lett. 105, 102602 (2014)
[5] X. Chen, P. Dai, D. Feng, T. Xiang, F.-C. Zhang, National Science Review 1 (3), 371- 395 (2014)
[6] J. Paglione, R. L. Greene, Nature Physics 6, 645 (2010)
[7] Y. Mizuguchi, Y. Takano, J. Phys. Soc. Jpn. 79, 102001 (2010)
[8] G. R. Stewart, Rev. Mod. Phys. 83,1589 (2011)
[9] A. V. Chubukov, P.J. Hirschfeld, Physics Today 68, 46 (2015)
[10] E.Dagotto, Rev. Mod. Phys. 85, 849 (2013)
[11] M.D.Lumsden, A.D.Christianson, E.A.Goremychkin, S.E.Nagler, H.A.Mook, M.B.Stone, D.L.Abernathy, T.Guidi, G.J.MacDougall, C. de la Cruz, A.S.Sefat, M.A.McGuire, B.C.Sales, D.Mandrus, Nature Physics 6, 182 (2010)
[12] C. Stock, E. E. Rodriguez, and M. A. Green Phys. Rev. B 85, 094507 (2012)
[13] M. Cutler, N.F. Mott, Phys. Rev. 181, 1336 (1969)
[14] R. R. Heikes, R. C. Miller, and R. Mazelsky, Physica 30, 1600 (1964)
[15] R. R. Heikes, in Thermoelectricity: Science and Engineering, edited by R. R. Heikes and R. W. Ure, Jr. (Interscience, New York, 1961).
[16] P.M.Chaikin, G.Beni, Physical Review B 13, 647 (1976)
[17] J.M. Ziman, Electrons and phonons (Oxford, Clarendon Press, 1960); R.D. Barnard, Thermoelectricity in metals and alloys (Taylor & Francis Ltd, London, 136, 1972).
[18] F. Caglieris, A. Braggio, I. Pallecchi, A. Provino, M. Pani, G. Lamura, A. Jost, U. Zeitler, E. Galleani D'Agliano, P. Manfrinetti, M. Putti, Phys,. Rev. B 90, 134421 (2014)

[19] G.N Grannemann, L. Berger, Phys. Rev. B 13, 2072 (1976)
[20] E. H. Sondheimer, Proc. R. Soc. London, Ser. A 193, 484 (1948)
[21] K. Behnia, J. Phys.: Condens. Matter 21, 113101 (2009)
[22] V. Oganesyan, I. Ussishkin, Phys. Rev. B 70, 054503 (2004)
[23] Y. Ran, F. Wang, H. Zhai, A. Vishwanath, D.-H. Lee, Phys. Rev. B 79, 014505 (2009).
[24] P. Richard, K. Nakayama, T. Sato, M. Neupane, Y.-M. Xu, J. H. Bowen, G. F. Chen, J. L. Luo, N. L. Wang, X. Dai, Z. Fang, H. Ding, T. Takahashi, Phys. Rev. Lett. 104, 137001 (2010)
[25] R.P. Huebener, "Magnetic flux structures in superconductors", Springer Series in Solid-State Sciences, vol. 6 (Springer, Berlin, 1979).
[26] R. P. Huebener, Supercond. Sci. Technol. 8, 189 (1995)
[27] K. Wang, H. Lei, C. Petrovic, Phys. Rev. B 83, 174503 (2011)
[28] Y. Wang, L. Li, N. P. Ong, Phys. Rev. B 73, 024510 (2006)
[29] A. Pourret, P. Spathis, H. Aubin, K. Behnia, New J. Phys. 11, 055071 (2009)
[30] A. A. Varlamov, A. V. Kavokin, Europhys. Lett. 103, 47005 (2013)
[31] M. A. McGuire, R. P. Hermann, A. S. Sefat, B. C. Sales, R. Jin, D. Mandrus, F. Grandjean, G. J Long, New J. Phys. 11, 025011 (2009)
[32] A.Kondrat, G.Behr, B.Büchner, C.Hess, Phys. Rev. B 83, 092507 (2011)
[33] A. Poddar, S. Mukherjee, T. Samanta, R. S. Saha, R. Mukherjee, P. Dasgupta, C. Mazumdar, R. Ranganathan, Physica C 469, 789 (2009)
[34] M. Matusiak, T. Plackowski, Z. Bukowski, N. D. Zhigadlo, J. Karpinski, Phys. Rev B 79, 212502 (2009)
[35] M. A. McGuire, A. D. Christianson, A. S. Sefat, B. C. Sales, M. D. Lumsden, R. Jin, E. A. Payzant, D. Mandrus, V. Varadarajan, J. W. Brill, R. P. Hermann, M. T. Sougrati, F. Grandjean, G. J. Long, Phys. Rev. B 78, 094517 (2008)
[36] M. Tropeano, I. Pallecchi, M. R. Cimberle, C. Ferdeghini, G. Lamura, M. Vignolo, A. Martinelli, A. Palenzona, M. Putti, Supercond. Sci. Technol. 23, 054001 (2010)
[37] Z.-W.Zhu, Q.Tao, Y.-K.Li, M.He, G.-H.Cao, Z.-A.Xu, Front. Phys. China 4(4), 455 (2009)
[38] B. Buchner, Private communication, 2012
[39] M. A. McGuire, D. J. Singh, A. S. Sefat, B. C. Sales, D. Mandrus, J. Sol. State Chem. 182, 2326 (2009)
[40] Y. Li, X. Lin, Q. Tao, C. Wang, T. Zhou, L. Li, Q. Wang, M. He, G. Cao, Z. Xu, New J. Phys. 11, 053008 (2009)
[41] M. Tropeano, M. R. Cimberle, C. Ferdeghini, G. Lamura, A. Martinelli, A. Palenzona, I. Pallecchi, A. Sala, I. Sheikin, F. Bernardini, M. Monni, S. Massidda, M. Putti, Phys. Rev. B 81, 184504 (2010)
[42] M. Tropeano, C. Fanciulli, F. Canepa, M. R. Cimberle, C. Ferdeghini, G. Lamura, A. Martinelli, M. Putti, M. Vignolo, A. Palenzona, Phys. Rev. B 79, 174523 (2009)
[43] T. M. McQueen, M. Regulacio, A. J. Williams, Q. Huang, J. W. Lynn, Y. S. Hor, D. V. West, M. A. Green, R. J. Cava, Phys. Rev. B 78, 024521 (2008)
[44] Q. Tao, Z. Zhu, X. Lin, G. Cao, Z. Xu, G. Chen, J. Luo, N. Wang, J. Phys.: Condens. Matter 22, 072201 (2010)
[45] S. Arsenijevic, R. Gaál, A. S. Sefat, M. A. McGuire, B. C. Sales, D. Mandrus, L. Forró, Phys. Rev. B 84, 075148 (2011)
[46] A. F. May, M. A. McGuire, J. E. Mitchell, A. S. Sefat, B. C. Sales, Phys. Rev. B 88, 064502 (2013)
[47] Y. J. Yan, X. F. Wang, R. H. Liu, H. Chen, Y. L. Xie, J. J. Ying, and X. H. Chen, Phys. Rev B 81, 235107 (2010)
[48] M.Matusiak, Z.Bukowski, J.Karpinski, Phys. Rev. B 81, 020510 (R) (2010)
[49] G. Wu, H. Chen, T. Wu, Y. L. Xie, Y. J. Yan, R. H. Liu, X. F. Wang, J. J. Ying, X. H. Chen, J. Phys.: Condens. Matter 20, 422201 (2008)
[50] M.Matusiak, Z.Bukowski, J.Karpinski, Phys. Rev. B 83, 224505 (2011)
[51] Z. Ren, Z. Zhu, S. Jiang, X. Xu, Q. Tao, C. Wang, C. Feng, G. Cao, Z. Xu, Phys. Rev. B 78, 052501 (2008)
[52] C. Liu, T. Kondo, R. M. Fernandes, A. D. Palczewski, E. D. Mun, N. Ni, A. N. Thaler, A. Bostwick, E. Rotenberg, J. Schmalian, S. L. Bud'ko, P. C. Canfield, and A. Kaminski, Nature Physics 6, 419 (2010)
[53] J. Maiwald, H. S. Jeevan, P. Gegenwart, Phys. Rev. B 85, 024511 (2012)
[54] S. Jiang, H. S. Jeevan, J. Dong, and P. Gegenwart, Phys. Rev. Lett. 110, 067001 (2013)
[55] J. H. Chu, J. G. Analytis, C. Kucharczyk, and I. R. Fisher, Phys. Rev. B 79, 014506 (2009)
[56] L. Zhang, D. J. Singh, M. H. Du, Phys. Rev. B 79, 012506 (2009)

[57] I. Pallecchi, G. Lamura, M. Tropeano, M. Putti, R. Viennois, E. Giannini, D. Van der Marel, Phys. Rev. B 80, 214511 (2009)

[58] M. Matusiak, E. Pomjakushina, K. Conder, Physica C 483, 21 (2012)

[59] S. Li, C. de la Cruz, Q. Huang, Y. Chen, J. W. Lynn, J. Hu, Y.-L. Huang, F.-C. Hsu, K.-W. Yeh, M.-K. Wu, P. Dai, Phys. Rev. B 79, 054503 (2009)

[60] W. Bao, Y. Qiu, Q. Huang, M. A. Green, P. Zajdel, M. R. Fitzsimmons, M. Zhernenkov, S. Chang, M. Fang, B. Qian, E. K. Vehstedt, J. Yang, H. M. Pham, L. Spinu, Z. Q. Mao, Phys. Rev. Lett. 102, 247001 (2009)

[61] A. Martinelli, A. Palenzona, M. Tropeano, C. Ferdeghini, M. Putti, M. R. Cimberle, T. D. Nguyen, M. Affronte, C. Ritter, Phys. Rev. B 81, 094115 (2010)

[62] F. Ma, W. Ji, J. Hu, Z.-Y. Lu, T. Xiang, Phys. Rev. Lett. 102, 177003 (2009)

[63] S. Chi, A. Schneidewind, J. Zhao, L. W. Harriger, L. Li, Y. Luo, G. Cao, Z. Xu, M. Loewenhaupt, J. Hu, P. Dai, Phys. Rev. Lett. 102, 107006 (2009)

[64] C. Zhang, H.-F. Li, Yu Song, Yixi Su, Guotai Tan, Tucker Netherton, Caleb Redding, Scott V. Carr, Oleg Sobolev, Astrid Schneidewind, Enrico Faulhaber, L. W. Harriger, Shiliang Li, Xingye Lu, Dao-Xin Yao, Tanmoy Das, A. V. Balatsky, Th. Brückel, J. W. Lynn, and Pengcheng Dai, Phys. Rev. B 88, 064504 (2013)

[65] C. Stock, E. E. Rodriguez, M. A. Green, P. Zavalij, J. A. Rodriguez-Rivera, Phys. Rev. B 84, 045124 (2011)

[66] H. Kotegawa, M. Fujita, Sci. Technol. Adv. Mater. 13, 054302 (2012)

[67] T. M. McQueen, Q. Huang, V. Ksenofontov, C. Felser, Q. Xu, H. Zandbergen, Y. S. Hor, J. Allred, A. J. Williams, D. Qu, J. Checkelsky, N. P. Ong, R. J. Cava, Phys. Rev. B 79, 014522 (2009)

[68] B. C. Sales, A. S. Sefat, M. A. McGuire, R. Y. Jin, D. Mandrus, Y. Mozharivskyj, Phys. Rev. B 79, 094521 (2009)

[69] F. Caglieris, F. Ricci, G. Lamura, A. Martinelli, A. Palenzona, I. Pallecchi, A. Sala, G. Profeta, M. Putti, Sci. Technol. Adv. Mater. 13, 054402 (2012)

[70] S. D. Obertelli, J. R. Cooper, J. L. Tallon, Phys. Rev. B 46, 14928(R) (1992)

[71] M. Putti, D. Marré, I. Pallecchi, P. G. Medaglia, A. Tebano, G. Balestrino, Phys. Rev. B 69, 134511 (2004)

[72] L. Pinsard-Gaudart, D. Berardan, J. Bobroff, N. Dragoe, Phys. Status Solidi (RRL) 2, 185 (2008)

[73] K. Kihou, C. H. Lee, K. Miyazawa, P. M. Shirage, A. Iyo, and H. Eisaki, J. Appl. Phys. 108, 033703 (2010)

[74] A. S. Sefat, M. A. McGuire, B. C. Sales, R. Jin, J. Y. Howe, D. Mandrus, Phys. Rev. B 77, 174503 (2008)

[75] N. Kaurav, Y. T. Chung, Y. K. Kuo, R. S. Liu, T. S. Chan, J. M. Chen, J.-F. Lee, H.-S. Sheu, X. L. Wang, S. X. Dou, S. I. Lee, Y. G. Shi, A. A. Belik, K. Yamaura, E. Takayama-Muromachi, Appl. Phys. Lett. 94, 192507 (2009)

[76] N. Kang, P. Auban-Senzier, C. R. Pasquier, Z. A. Ren, J. Yang, G. C. Che, Z. X. Zhao, New J. Phys. 11, 025006 (2009)

[77] J. Prakash, S. J. Singh, S. Patnaik, A. K. Ganguli, J. Phys.: Condens. Matter 21, 175705 (2009)

[78] L.-J. Li, Y.-K. Li, Z. Ren, Y.-K. Luo, X. Lin, M. He, Q. Tao, Z.-W. Zhu, G.-H. Cao, Z.-A. Xu, Phys. Rev. B 78, 132506 (2008)

[79] S. J. Singh, J. Prakash, S. Patnaik, A. K. Ganguli, Supercond. Sci. Technol. 22, 045017 (2009)

[80] A. S. Sefat, A. Huq, M. A. McGuire, R. Jin, B. C. Sales, D. Mandrus, L. M. D. Cranswick, P. W. Stephens, K. H. Stone, Phys. Rev. B 78, 104505 (2008)

[81] C. Wang, Y. K. Li, Z. W. Zhu, S. Jiang, X. Lin, Y. K. Luo, S. Chi, L. J. Li, Z. Ren, M. He, H. Chen, Y. T. Wang, Q. Tao, G. H. Cao, Z. A. Xu, Phys. Rev. B 79, 054521 (2009)

[82] L.-D. Zhao, D. Berardan, C. Byl, L. Pinsard-Gaudart, N. Dragoe, J. Phys.: Condens. Matter 22, 115701 (2010)

[83] Y. K. Li, X. Lin, T. Zhou, J. Q. Shen, Q. Tao, G. H. Cao, Z. A. Xu, J. Phys.: Condens. Matter 21, 355702 (2009)

[84] X. Lin, H. J. Guo, C. Y. Shen, Y. K. Luo, Q. Tao, G. H. Cao, Z. A. Xu, Phys. Rev. B 83, 014503 (2011)

[85] Y. K. Li, X. Lin, Q. Tao, H. Chen, C.Wang, L. J. Li, Y. K. Luo, M. He, Z. W. Zhu, G. H. Cao, Z. A. Xu, Chin. Phys. Lett. 26, 017402 (2009)

[86] S. C. Lee, E. Satomi, Y. Kobayashi, M. Sato, J. Phys. Soc. Jpn. 79, 023702 (2010)

[87] G. Xu, W. Ming, Y. Yao, X. Dai, S. C. Zhang, Z. Fang, Europhys. Lett. 82, 67002 (2008)

[88] M. Sato, Y. Kobayashi, S. C. Lee, H. Takahashi, E. Satomi, Y. Miura, J. Phys. Soc. Jpn. 79, 014710 (2010)

[89] S.J. Singh, J. Shimoyama, A. Yamamoto, H. Ogino, K. Kishio, Physica C 494, 57 (2013)

[90] S. J. Singh, J.-I. Shimoyama, A. Yamamoto, H. Ogino, K. Kishio, Physica C 504, 19 (2014)

[91] H. M. Hodovanets, A. Thaler, E. Mun, N. Ni, S. L. Bud'ko, P. C. Canfield, Phil. Mag. 93, 661 (2013)

[92] E. D. Mun, S. L. Bud'ko, N. Ni, A. N. Thaler, P. C. Canfield, Phys. Rev. B 80, 054517 (2009)

[93] H. Hodovanets, E. D. Mun, A. Thaler, S. L. Bud'ko, P. C. Canfield, Phys. Rev. B 83, 094508 (2011)

[94] B.C. Sales, M. A. McGuire, A. S. Sefat, D. Mandrus, Physica C 470, 304 (2010)

[95] S. Arsenijevic, H. Hodovanets, R. Gaál, L. Forró, 1 S. L. Bud'ko, P. C. Canfield, Phys. Rev. B 87, 224508 (2013)

[96] M. Gooch, B. Lv, B. Lorenz, A. M. Guloy, C.-W. Chu, Phys. Rev B 79, 104504 (2009)

[97] A. Thaler, H. Hodovanets, M. S. Torikachvili, S. Ran, A. Kracher, W. Straszheim, J. Q. Yan, E. Mun, P. C. Canfield, Phys. Rev. B 84, 144528 (2011)

[98] B. Lv, M. Gooch, B. Lorenz, F. Chen, A. M. Guloy, C. W. Chu, New J. Phys. 11, 025013 (2009)

[99] H. Hodovanets, Y. Liu, A. Jesche, S. Ran, E. D. Mun, T. A. Lograsso, S. L. Bud'ko, P. C. Canfield, Phys. Rev. B 89, 224517 (2014)

[100] G. Wu, R. H. Liu, H. Chen, Y. J. Yan, T. Wu, Y. L. Xie, J. J. Ying, X. F. Wang, D. F. Fang, X. H. Chen, Europhys. Lett. 84 27010 (2008)

[101] W. Malaeb, T. Shimojima, Y. Ishida, K. Okazaki, Y. Ota, K. Ohgushi, K. Kihou, T. Saito, C. H. Lee, S. Ishida, M. Nakajima, S. Uchida, H. Fukazawa, Y. Kohori, A. Iyo, H. Eisaki, C.-T. Chen, S. Watanabe, H. Ikeda, and S. Shin, Phys. Rev. B 86, 165117 (2012)

[102] N. Xu, P. Richard, X. Shi, A. van Roekeghem, T. Qian, E. Razzoli, E. Rienks, G.-F. Chen, E. Ieki, K. Nakayama, T. Sato, T. Takahashi, M. Shi, and H. Ding, Phys. Rev. B 88, 220508 (2013)

[103] Y. Liu, T. A. Lograsso, Phys. Rev. B 90, 224508 (2014)

[104] S. N. Khan, D. D. Johnson, Phys. Rev. Lett. 112, 156401 (2014)

[105] S. Thirupathaiah, E. D. L. Rienks, H. S. Jeevan, R. Ovsyannikov, E. Slooten, J. Kaas, E. van Heumen, S. de Jong, H. A. Dürr, K. Siemensmeyer, R. Follath, P. Gegenwart, M. S. Golden, J. Fink, Phys. Rev. B 84, 014531 (2011)

[106] C.-H. Wang, T.-K. Chen, C.-C. Chang, C.-H. Hsu, Y.-C. Lee, M.-J. Wang, P. M. Wu, M.-K. Wu, Europhys. Lett. 111, 27004 (2015)

[107] T.-K. Chen, C.-C. Chang, H.-H. Chang, A.-H. Fang, C.-H. Wang, W.-H. Chao, C.-M. Tseng, Y.-C. Lee, Y.-R. Wu, M.-H. Wen, H.-Y. Tang, F.-R. Chen, M.-J. Wang, M.-K. Wu, D. Van Dyck, PNAS 111, 63 (2014)

[108] A. Chubukov, Ann. Rev. Cond. Matt. Phys. 3, 57 (2012)

[109] Y. J. Yan, M. Zhang, A. F. Wang, J. J. Ying, Z. Y. Li, W. Qin, X. G. Luo, J. Q. Li, Jiangping Hu, X. H. Chen, Sci. Rep. 2, 212 (2012)

[110] A. Pourret, L. Malone, A. B. Antunes, C. S. Yadav, P. L. Paulose, B. Fauqué, K. Behnia, Phys. Rev. B 83, 020504(R) (2011)

[111] R. Hu, K. Cho, H. Kim, H. Hodovanets, W. E. Straszheim, M. A. Tanatar, R. Prozorov, S. L. Bud'ko, P. C. Canfield, Supercond. Sci. Technol. 24, 065006 (2011)

[112] S. Margadonna, Y. Takabayashi, M. T. McDonald, K. Kasperkiewicz, Y. Mizuguchi, Y. Takano, A. N. Fitch, E. Suarde, K. Prassides, Chem. Commun. 5607-5609 (2008)

[113] R. Khasanov, M. Bendele, K. Conder, H. Keller, E. Pomjakushina, V. Pomjakushin, New J. Phys. 12, 073024 (2010)

[114] E. Pomjakushina, K. Conder, V. Pomjakushin, M. Bendele, R. Khasanov, Phys. Rev. B 80, 024517 (2009)

[115] Y. J. Song , J. B. Hong , B. H. Min, Y. S. Kwon, K. J. Lee, M. H. Jung, J.-S. Rhyee, Journal of the Korean Physical Society 59, 312 (2011)

[116] E. L. Thomas, W. Wong-Ng, D. Phelan, J. N. Millican, J. Appl. Phys. 105, 073906 (2009)

[117] C.-J. Liu, A. Bhaskar, H.-J. Huang, F.-H. Lin, Appl. Phys. Lett. 104, 252602 (2014)

[118] M. A. Tanatar, J.-Ph. Reid, S. René de Cotret, N. Doiron-Leyraud, F. Laliberté, E. Hassinger, J. Chang, H. Kim, K. Cho, Y. J. Song, Y. S. Kwon, R. Prozorov, L. Taillefer, Phys. Rev. 84, 054507 (2011)

[119] C. W. Chu, F. Chen, M. Gooch, A. M. Guloy, B. Lorenz, B. Lv, K. Sasmal, Z. J. Tang, J. H. Tapp, and Y. Y. Xue, Physica C 469, 326 (2009)

[120] X. F. Xu, Z. A. Xu, T. J. Liu, D. Fobes, Z. Q. Mao, J. L. Luo, Y. Liu, Phys. Rev. Lett. 101, 057002 (2008)

[121] I. Pallecchi, F. Bernardini, M. Tropeano, A. Palenzona, A. Martinelli, C. Ferdeghini, M. Vignolo, S. Massidda, M. Putti, Phys. Rev. B 84, 134524 (2012)

[122] I. Pallecchi, F.Bernardini, F.Caglieris, A.Palenzona, S.Massidda, M.Putti, Eur. Phys. J. B 86, 338 (2013)

[123] K. K. Huynh, Y. Tanabe, K. Tanigaki, Phys. Rev. Lett. 106, 217004 (2011)

[124] Y. Kim, H. Oh, C. Kim, D. Song, W. Jung, B. Kim, H. J. Choi, C. Kim, B. Lee, S. Khim, H. Kim, K. Kim, J. Hong, Y. Kwon, Phys. Rev. B 83, 064509 (2011)

[125] T. Morinari, E. Kaneshita, T. Tohyama, Phys. Rev. Lett. 105, 037203 (2010)

[126] Z. W. Zhu, Z. A. Xu, X. Lin, G. H. Cao, C. M. Feng, G. F. Chen, Z. Li, J. L. Luo, N. L. Wang, New J. Phys. 10, 063021 (2008)

[127] C. Hess, "Properties and applications of thermoelectric materials – II", V. Zlatic and A. Hewson editors, Proceedings of NATO Advanced Research Workshop, Hvar, Croatia, September 19-25, 2011, NATO Science for Peace and Security Series B: Physics and Biophysics (Springer Science+Business Media B.V. 2012), arXiv:1202.2959

[128] A. Hackl, M. Vojta, S. Sachdev, Phys. Rev. B 81, 045102 (2010).

[129] I. Pallecchi, M. Tropeano, G. Lamura, M. Pani, M. Palombo, A. Palenzona, M. Putti, Physica C 482, 68 (2012)

[130] K. K. Huynh, Y. Tanabe, T. Urata, H. Oguro, S. Heguri, K. Watanabe, K. Tanigaki, Phys. Rev. B 90, 144516 (2014)

[131] F. Ricci, G. Profeta, Phys. Rev. B 87, 184105 (2013)

[132] E. Bellingeri, S. Kawale, F. Caglieris, V. Braccini, G. Lamura, L. Pellegrino, A. Sala, M. Putti, C. Ferdeghini, A. Jost, U. Zeitler, C. Tarantini, J. Jaroszynski, Supercond. Sci. Technol. 27, 044007 (2014)

[133] I. Pallecchi, C. Fanciulli, M. Tropeano, A. Palenzona, M. Ferretti, A. Malagoli, A. Martinelli, I. Sheikin, M. Putti, and C. Ferdeghini, Phys. Rev. B 79, 104515 (2009)